\documentclass[]{pasj01}
\draft

\Received{}%{yyyy/mm/dd}
\Accepted{}%{yyyy/mm/dd}
\usepackage{graphicx}	% Including figure files
\usepackage{rotating}
\usepackage{txfonts}
\usepackage{tabularx}

\usepackage{footnote}
\begin{document} 

\title{ 
%\LETTERLABEL %%% <-- uncomment for LETTER article  
%\REVIEWLABEL %%% <-- uncomment for REVIEW article  
Searching for Hot Subdwarf Stars in LAMOST DR1 - II.
Pure spectroscopic identification method for  hot subdwarfs}

%%% begin:list of authors
% Do NOT capitalize all letters in "textsc".
\author{Zhenxin \textsc{Lei}\altaffilmark{1,2}}
\altaffiltext{1}{Key Laboratory of Optical Astronomy, National Astronomical Observatories, Chinese Academy of Sciences, Beijing 100012, China}
\altaffiltext{2}{College of Science, Shaoyang University, Shaoyang 422000, China}
\email{zxlei@nao.cas.cn}

\author{Yude \textsc{Bu},\altaffilmark{3}}
\altaffiltext{3}{School of Mathematics and Statistics, Shandong University, Weihai, 264209, Shandong, China}
%\email{bbbbb@xxx.xxx.xx.xx}

\author{Jingkun \textsc{Zhao}\altaffilmark{1}}
%\altaffiltext{1}{C-Address of Institute}
%\email{ccccc@xxx.xxx.xx.xx}

\author{P\'eter \textsc{N\'emeth},\altaffilmark{4,5}}
\altaffiltext{4}{Astronomical Institute of the Czech Academy of Sciences, CZ-251\,65, Ond\v{r}ejov, Czech Republic}
\altaffiltext{5}{Astroserver.org, 8533 Malomsok, Hungary}
%\email{bbbbb@xxx.xxx.xx.xx}

\author{Gang \textsc{Zhao}\altaffilmark{1}}
%\altaffiltext{1}{C-Address of Institute}
\email{gzhao@nao.cas.cn}
%%% end:list of authors

%% `\KeyWords{}' always has to be placed before ``\maketitle'' 
%%  List of Key Words:  https://academic.oup.com/pasj/pages/Pasj_Keywords 

\maketitle

\begin{abstract}
Employing a new machine learning method, named hierarchical extreme learning machine (HELM) algorithm, 
we identified 56 hot subdwarf stars in the first data release (DR1) of 
the Large Sky Area Multi-Object Fibre Spectroscopic Telescope (LAMOST) survey. 
The atmospheric parameters of the stars are obtained 
by fitting the profiles of hydrogen (H) Balmer lines and helium (He) lines with synthetic spectra calculated from 
non-Local Thermodynamic Equilibrium (NLTE)  
model atmospheres. Five He-rich hot subdwarf stars were found in our sample with their 
$\mathrm{log}(n\mathrm{He}/n\mathrm{H})>-1$, while 51  
stars are He-poor sdB,  sdO and sdOB stars. We also confirmed the two 
He sequences of hot subdwarf stars found by Edelmann et al. (2003) 
in $T_\mathrm{eff}$-$\mathrm{log}(n\mathrm{He}/n\mathrm{H})$ diagram. 
The HELM algorithm works directly on the observed spectroscopy 
and is able to filter out spectral properties without supplementary photometric data.
The results presented  in this study demonstrate that the HELM algorithm 
is a reliable method to search for 
hot subdwarf stars after a suitable training is performed, and it is also 
suitable to  search for other objects which have obvious 
features in their spectra or images. 
\end{abstract}

\section{Introduction}

Hot subdwarf stars (spectral types i.e.: sdB,  sdO and related objects) are low mass 
stars in a core or shell helium (He) burning stage (Heber 2009, 2016).  These stars 
lose nearly their whole hydrogen (H) envelopes during the evolution 
on the red giant branch (RGB), therefore they present very  high effective temperatures 
($T_\mathrm{eff}$ $\geq$ 20 000 K) on reaching the horizontal branch (HB) stage. 
Hot subdwarf stars are considered to be the main source of UV-excess 
found in elliptical galaxies (O'Connell 1999; Han et al. 2007). These stars 
also turned out to be important objects in studying close binary interactions, since many 
hot subdwarf  stars are found in close binaries (Maxted et al. 2001; Napiwotzki et al. 2004; 
Copperwheat et al. 2011). The most common types of  
companion stars in hot subdwarf binaries are 
main-sequence (MS) stars, white dwarfs (WDs), brown dwarfs and planets. Hot 
subdwarf stars with massive WD companions are considered to be the progenitors 
of type Ia supernovae (Wang et al. 2009; Geier et al. 2011; Geier 2015). The atmospheres of hot subdwarf 
stars are good places to study diffusion processes, such as gravitational settling and 
radiative levitation. Moreover, pulsating sdB/O stars are extensively used in  
asteroseismology to study stellar interiors and rotation. 
For a recent review on hot subdwarf stars see Heber 2016. 

The formation mechanism of hot subdwarf stars is still unclear. Since  about half of the 
hot subdwarf B type (sdB) stars are found in close binaries, Han et al. (2002, 2003) 
carried out a detailed binary population synthesis to study the formation of sdB stars. 
They found that common envelope (CE) ejection, mass transfer through Roche lobe overflow (RLOF) 
or merger of two helium core white dwarfs (He-WDs) could produce sdB stars in a close binary, 
wide binary and single system respectively. Based on these results, Chen et al. (2013) 
predicted that the orbital period of sdB binaries formed from RLOF mass transfer 
could be up to 1200 days, if atmospheric RLOF and a different angular momentum 
loss  are considered in binary evolution. This result could explain the formation of sdB stars 
found in wide binaries. Furthermore, Xiong et al. (2017) found that two distinct  groups of sdB stars 
could be formed through the detailed  CE ejection channel. One group  
is flash-mixing sdB stars without H-rich envelopes, and the other is canonical 
sdB stars with H-rich envelopes. In addition, Zhang et al. (2012, 2017) studied the 
formation channel in detail for single sdB stars through the merger of two He-WDs or the merger of  a He-WD 
with a low-mass MS companion. Their results could account for some He-rich sdB stars found 
in the field. The counterpart of hot sudwarf stars in globular clusters (GCs) 
are known as extreme horizontal branch (EHB) stars. Some of these stars with particularly  high 
effective temperatures (e.g., $T_\mathrm{eff}$ $\geq$ 32 000 K ) form a blue hook in the ultraviolet (UV) 
color-magnitude diagram (CMD) of GCs (Brown et al. 2016), and they are 
known as blue hook stars in GCs. Lei et al. (2015, 2016) proposed that 
tidally-enhanced stellar wind during binary evolution may lead to huge mass 
loss of the primary stars at RGB and could produce blue hook stars in GCs after 
undergoing late core He flash. 

Thanks to large surveys over the past decade a significant 
number of previously unknown hot subdwarfs  have been catalogued, e.g., 
Kepler ($\varnothing$stensen et al. 2010), Galaxy Evolution Explorer 
(GALEX, Vennes et al. 2011; N\'emeth et al. 2012; Kawka et al. 2015), the Sloan Digital Sky Survey 
(SDSS, Geier et al. 2015; Kepler et al. 2015, 2016) 
and the Large Sky Area Multi-Object Fibre Spectroscopic Telescope (LAMOST) survey 
(Luo et al. 2016). $\varnothing$stensen (2006) compiled a widely used hot subdwarf 
database by searching extensive literatures, in which more than 2300 hot subdwarf stars 
are archived. Furthermore, Geier et al. (2017) compiled a  catalogue of known hot subdwarf 
stars and candidates retrieved from literatures and unpublished databases. This catalogue 
contains 5613 objects with multi-band photometry, proper 
motions, classifications, atmospheric parameters, radial velocities and 
information on light curve variability. Using the first data release (DR1) of the LAMOST survey, Luo et al. (2016) 
identified 166 hot subdwarf stars, among which 122 objects are single-lined, while the 
other 44 objects present double-lined composite spectra (e.g., Mg I triplet lines at 5170 $\mathrm{\AA}$ 
or Ca II triplet lines at 8650 $\mathrm{\AA}$) , which demonstrates the binary nature of these stars. 

We need even more spectroscopically identified hot subdwarf stars and
candidates to improve our understanding on their formation and evolution. Fortunately, 
large spectroscopic surveys provide us a good opportunity to 
search for new hot subdwarf stars, e.g., SDSS (York et al. 2000) and LAMOST 
(Cui et al. 2012; Zhao et al. 2006, 2012). The traditional method extensively used 
to search for hot subdwarf stars in large spectroscopic surveys 
is based on color cuts, followed by visual inspections.  However, this method 
requires homogeneous photometry for the spectra to 
obtain their colors in different band (e.g., \textit{u-g} and \textit{g-r}, Geier et al. 2011), thus it might  
not work well in spectral database without any or lack of  homogeneous photometric 
information, such as the database of LAMOST. 

Employing the Hierarchical Extreme Learning 
Machine (HELM) algorithm, Bu et al. (2017, hereafter Paper I) 
explored a machine learning method to search for hot subdwarf stars in LAMOST spectra. 
The Extreme Learning Machine (ELM)  is a special type of single hidden-layer feed-forward network, 
while HELM is the hierarchical framework of the ELM algorithm (Huang et al. 2006). 
It is inspired by the deep learning algorithms, and built in a multilayer manner. HELM 
has been frequently used in many fields, such as image-quality assessment (Mao et al. 2014), 
human action recognition (Minhas et al. 2010) and hyper-spectral image classification 
(Li et al. 2015). Using the HELM algorithm in Paper I, we obtained an  
accuracy and efficiency of classifying single-lined hot subdwarf stars in LAMOST spectra up to 
92\% and 96\% respectively, which demonstrated the reliability of the method to 
search for hot subdwarf stars in  the LAMOST survey spectral database. 

Like in the seminal study of Paper I, we applied the 
HELM algorithm method to LAMOST  DR1 and 
identified 56 hot subdwarf stars. We  obtained the atmospheric parameters of these stars 
by fitting their spectra with synthetic spectra calculated from  
NLTE model atmospheres (N\'emeth et al. 2012, 2014). The structure of the paper 
is as follows. In Section 2, we briefly introduced the LAMOST spectral survey and 
sample filtering method based on the HELM algorithm.  
In Section 3, we introduced the selection criteria to sort out hot 
subdwarf stars selected from the candidates by the HELM algorithm. We give our results in Section 4. Finally, 
a discussion and a summary of this study are presented in Section 5 and 6, respectively. 

\section{The Lamost survey and sample filtering with the HELM algorithm} \label{sec:LAMOST and sample}
\subsection{The LAMOST survey and database DR1}
LAMOST is a special reflecting Schmidt telescope designed with both large aperture 
(effective aperture of 3.6 - 4.9 m) and a wide field of view (FOV, 5$^{\circ}$, 
Cui et al. 2012). LAMOST is equipped with 16 low resolution spectrographs connected to 
4000 optical fibres, which are precisely positioned on the focal surface. 
As the telescope with the highest rate of spectral acquisition all over the world, LAMOST could 
obtain the spectra of 4000 objects simultaneously.  

LAMOST conducted its pilot survey between October 2011 and June 2012, 
while the regular survey started in September 2012 and finished its 
first year's operation in June 2013. The data from both the pilot survey and 
the first year regular survey make up the database of  LAMOST DR1 (Luo et al. 2015). 
DR1 contains totally 2\,204\,696 spectra with a resolution ($\lambda/\Delta\lambda$) of 1800 in the 
wavelength range 3690-9100$\mathrm{\AA}$, among which 
1\,790\,879 spectra have their signal-to-noise ratio (SNR) $\geq$10, and 
1\,944\,329 spectra are classified as stellar spectra. Although the 
number of stellar spectra in LAMOST DR1 is  large, many of them 
lack  photometric measurements in certain  bands, such as the \textit{u} band, and it  
prevents one to use colors for object classifications. 
Therefore, LAMOST DR1 provides us an appropriate database 
to test our new method (HELM algorithm) in searching for 
hot subdwarf stars directly from observed spectra, without a need for color information   
(also see the discussion in Section 5). 

\subsection{The HELM algorithm and our training sample}
HELM stands for the hierarchical framework of the ELM algorithm 
(see Paper I for more details), which was proposed by Tang et al. (2015). It 
usually contains two parts: an unsupervised learning part and  a supervised part. 
The unsupervised part in HELM could include many layers. To give higher-level features 
of the training sample, the input of each layer is  the output of the previous layer. On 
the other hand, the supervised part contains only one layer,  and 
it takes the output of the last unsupervised layer as its input. In the experiments of  Paper I, The HELM algorithm 
could filter out single-lined hot subdwarf stars from LAMOST spectra with an accuracy of 0.92 and  efficiency of 0.96,  respectively. When applied to the selection of double-lined hot subdwarfs, 
the HELM presented an accuracy and efficiency of 0.80 and 0.71, respectively. 
These results are better when we compare them with other popular algorithms (see section 4.2 in Paper I), 
which demonstrates that the HELM algorithm is an accurate and efficient new method 
to search for hot subdwarf stars in large spectroscopic surveys. 

The training sample used in the experiments of Paper I are the spectra of hot subdwarf stars identified in 
Luo et al. (2016) combined with 4600 LAMOST DR1 spectra of various types of objects, 
including stars of different spectral types, galaxies, 
quasars and objects with ambiguous spectral features. There are  a total of 166 hot subdwarf spectra 
in our training sample, among which 
122 stars are single-lined hot subdwarfs, while 44 spectra show strong Mg I triplet lines at 5170 $\mathrm{\AA}$ 
or Ca II triplet lines at 8650 $\mathrm{\AA}$ indicating the binary nature of these stars. 
According to Table 2 in Luo et al. (2016), the 122 single-lined hot subdwarf stars consist of 77 
sdB stars, 15 He-sdO stars, 12 sdO stars, 10 He-sdB stars and 8 blue horizontal branch (BHB) stars. 
All the sample spectra are divided into three groups to carry out the experiments in HELM and other 
popular algorithms (see Paper I for details). 

\section{Target selection}
\begin{figure*}
\centering
\begin{minipage}[c]{0.3\textwidth}
\includegraphics[width=45mm]{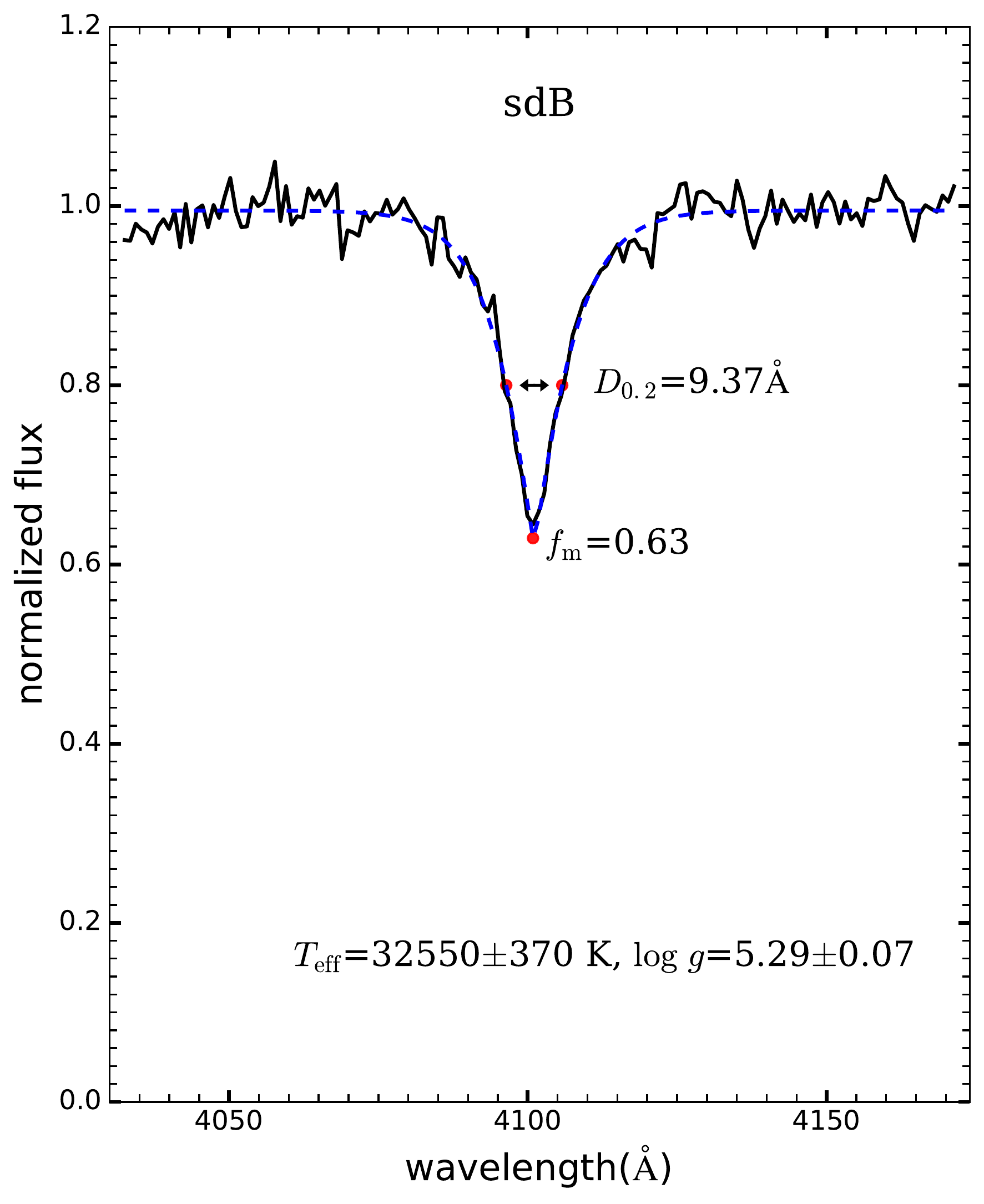}
\centerline{(a)}
\end{minipage}%
\centering
\begin{minipage}[c]{0.3\textwidth}
\includegraphics[width=45mm]{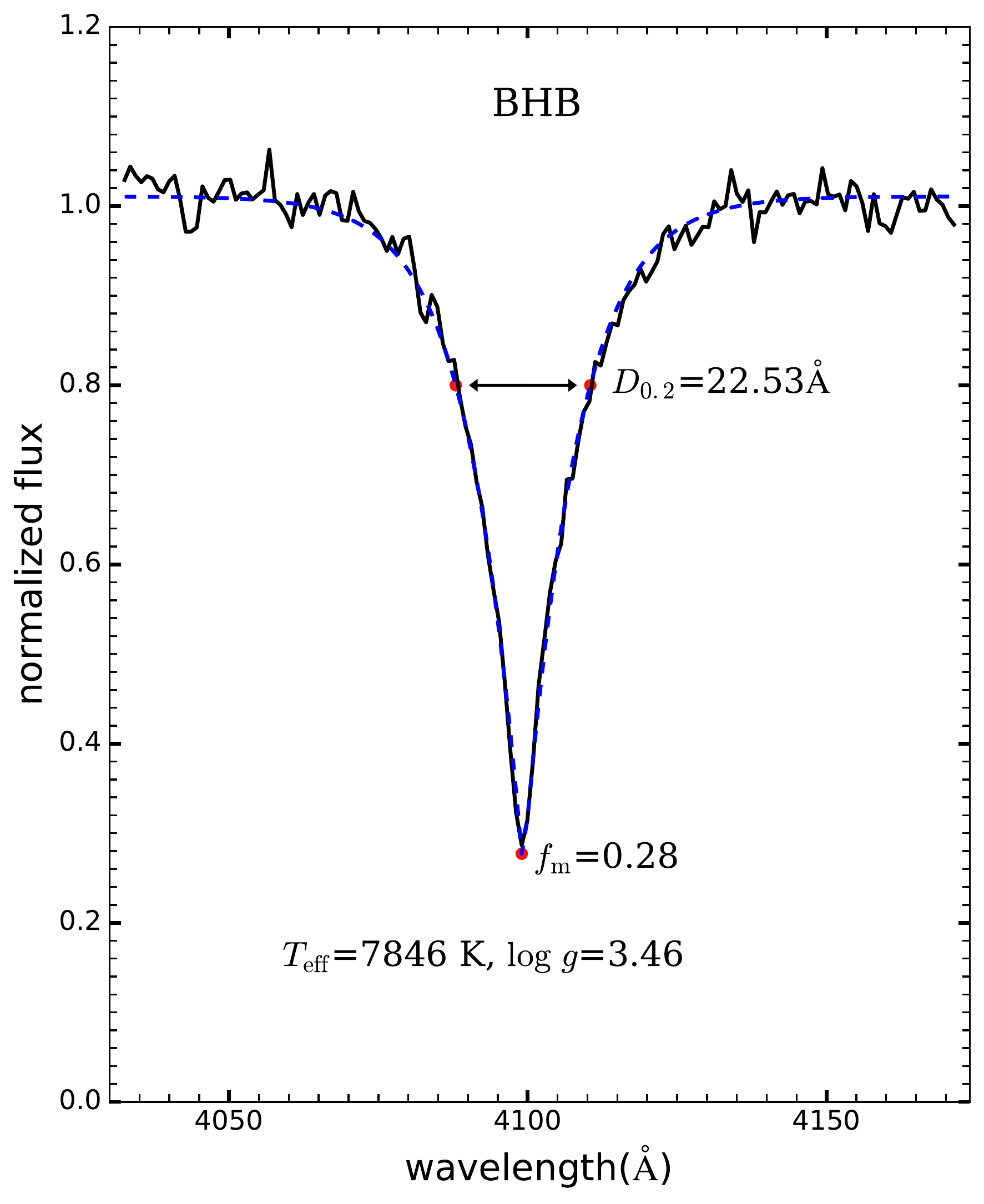}
\centerline{(b)}
\end{minipage}%
\centering
\begin{minipage}[c]{0.3\textwidth}
\includegraphics[width=45mm]{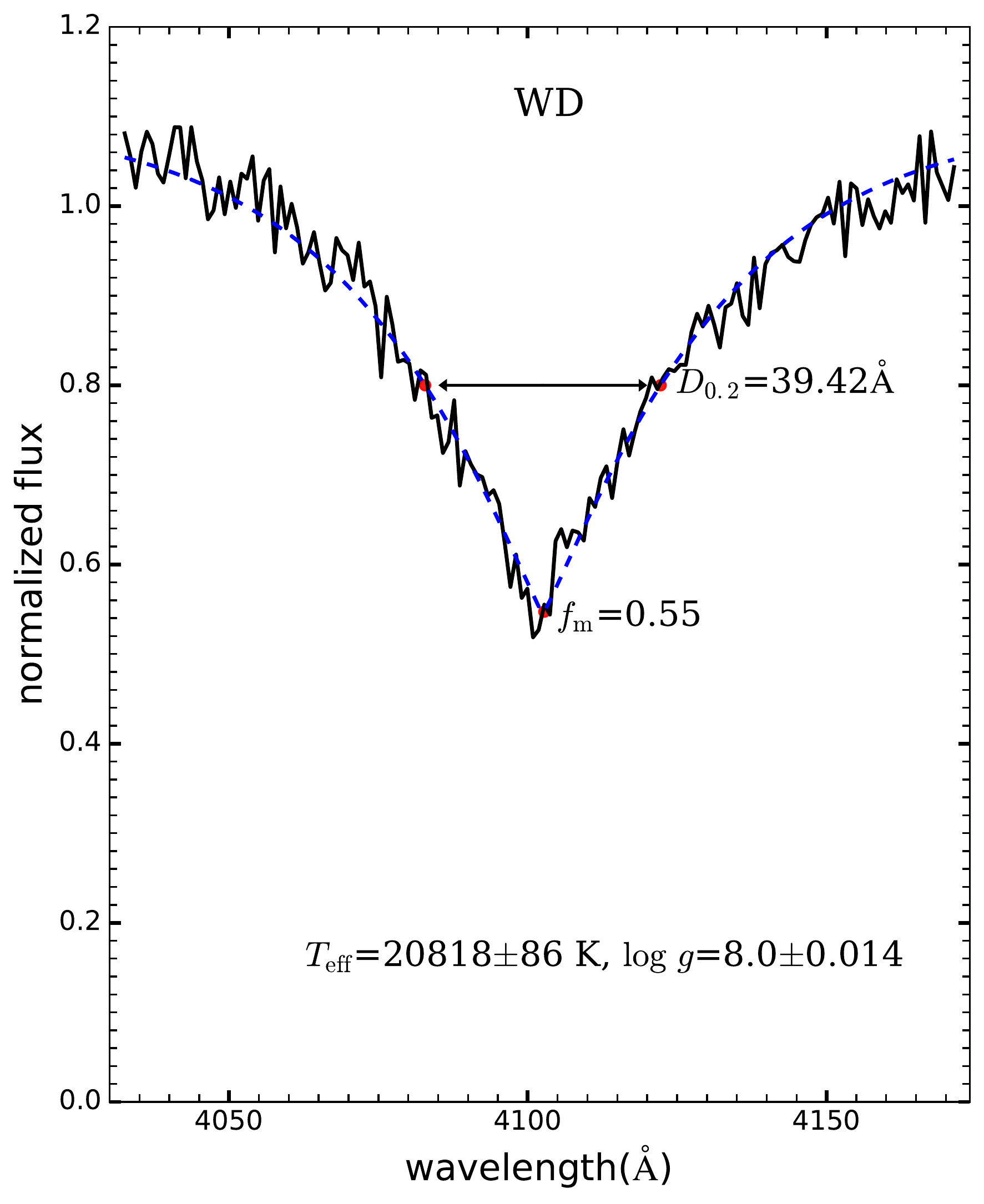}
\centerline{(c)}
\end{minipage}

 %\begin{minipage}[]{200mm}
\caption{Normalized spectra near the H$_\delta$ line in three different types of stars. 
The blue dashed curve is the fitting profile of $\mathrm{H}_{\delta}$ line. 
The values of  $D_{0.2}$ and $f_\mathrm{m}$ for each star are showed.}
%\end{minipage}
\end{figure*}

By applying the HELM algorithm outlined in Paper I, we obtained more than 
7000 hot subdwarf candidates from LAMOST DR1, among which 1034 
spectra have an $u$-band SNR larger than 10. We have selected our final 
hot subdwarf sample from these candidates. 
Blue horizontal branch (BHB) stars, B-type main-sequence (B-MS) stars and 
WDs show very similar features (e.g., strong H Balmer lines) in 
their spectra as  hot subdwarf stars  
(Moehler et al. 1990). Some of 
these stars have similar temperatures to hot subdwarf stars, especially to  
He-poor sdB stars. Therefore, the hot subdwarf candidate sample     
selected by the HELM algorithm method is contaminated by the above mentioned object types.  
Three steps are used to select hot subdwarf stars from our candidates. 

\subsection{Excluding BHB stars  and WDs from our sample}
BHB stars are horizontal branch stars bluer than the  RR Lyrae instability strip 
in the color-magnitude diagram (CMD). These stars present effective temperatures 
in the range of about 7 000 - 20 000 K and surface gravities (e.g., log\ $g$) in the range of 
$\log\ {g}\ =\ 2.5-4.5$ cm\,s$^{-2}$, respectively (Catelan 2009). Xue et al. (2008) 
used the $D_{0.2}$ and $f_\mathrm{m}$ method to discriminate BHB stars from 
blue straggler (BS)  and  B-MS stars. In this method, $D_{0.2}$ is the full 
width of the H$_\delta$  line at 20\% below the local continuum, while $f_\mathrm{m}$ 
is the flux relative to the continuum at the line core (Beers et al. 1992; Sirko et al. 2004). 
Xue et al. (2008) used the criteria:  $17\mathrm {\AA} \leq D_{0.2} \leq 28.5 \mathrm {\AA} $
and $0.1 \leq f_\mathrm{m}\leq 0.3$, to select BHB stars from their samples. 

Both the values of $D_{0.2}$ and $f_\mathrm{m}$ are sensitive to 
effective temperature and gravity in hot stars (Xue et al. 2008), which makes it 
a suitable measure to  distinguish our sample spectra in the $D_{\rm 0.2}$ - $f_{\rm m}$ diagram. 
Since BHB stars have lower temperatures and gravities than 
hot subdwarf stars and regular WDs present higher temperatures and 
gravities than hot subdwarf stars, these spectral classes can be 
clearly separated according to  their $D_{\rm 0.2}$ and $f_{\rm m}$ values 
(Greenstein \& Sargent 1974). We use 
\textit{the scale width versus shape method} (Clewley et al. 2002; Xue et al. 2008) to fit the 
$\mathrm{H}_{\delta}$ line  and obtain the value of $D_{0.2}$ and $f_\mathrm{m}$ for 
each spectrum in our sample. This method is based on 
a S\'ersic  profile fit (S\'ersic 1968) to Balmer lines in the following form: 
\begin{equation}
y=1.0-a \; exp\left[-\left(\frac{|\lambda-\lambda_{0}|}{b}\right)^{c}\right], 
\end{equation}
where $y$ is the normalized flux, $\lambda$ is the wavelength and $\lambda_0$ 
is the nominal central wavelength of the Balmer line.  
The coefficients $a$, $b$ and $c$ are free parameters.   
As described in Xue et al. (2008), to account for 
imperfect normalization of spectra, we used five free parameters: $a$, $b$, $c$, $\lambda_0$ and $n$ to fit the normalized spectrum to the S\'ersic profile: 
\begin{equation}
y=n-a \; exp\left[-\left(\frac{|\lambda-\lambda_{0}|}{b}\right)^{c}\right]. 
\end{equation}

The three panels in Fig 1 show the results of fitting the $\mathrm{H}_{\delta}$ profile of a 
sdB star, a BHB star and a  WD,  respectively. In each panel, 
solid curves  represent an extracted spectrum near the  
$\mathrm{H}_{\delta}$ line, while blue dashed curves denote our best fitting line profiles.   
Panel (a)  shows the spectrum of the sdB star 
PG\,1605+072 taken from Luo et al. (2016) 
with $T_\mathrm{eff}$\ =\ 32\,550$\pm$370 K 
and $\mathrm{log}\ g$=\ 5.29$\pm$0.07 cm\,s$^{-2}$. By adopting the 
fitting method described above, we got $D_{0.2}$\ =\ 9.37 $\mathrm{\AA}$ and $f_\mathrm{m}$\ 
=\ 0.63.  Panel (b) shows the spectrum of the BHB 
star SDSSJ171935.27+262234.9 from Xue et al. (2008) with $T_\mathrm{eff}$ = 7846 K 
and $\mathrm{log}\ g$\ =\ 3.46 cm\,s$^{-2}$ (no error bars for this star are presented in Xue et al. 2008 ), 
while its $D_{0.2}$ and $f_\mathrm{m}$ are 22.53 $\mathrm{\AA}$ and 0.28, respectively. 
One can see obviously that the BHB star presents much deeper 
$\mathrm{H}_{\delta}$ line (i.e., smaller value of $f_\mathrm{m}$) 
and  much wider $D_{0.2}$ than the sdB star in Panel (a) due to 
its significantly lower effective temperature and gravity. 
The spectrum of the WD SDSS\,J094126.79+294503.4 in Panel (c) is taken from the catalogue of 
Eisenstein et al. (2006) with its $T_\mathrm{eff}$ =\ 20\,818 K and $\mathrm{log} \ g$\ =\ 8.0 cm\,s$^{-2}$.  
Although this WD shows a similar depth of the  $\mathrm{H}_{\delta}$ line (i.e., $f_\mathrm{m}$\ =\ 0.55) 
to the sdB star showed in Panel (a), it presents a much larger $D_{0.2}$ (39.42 $\mathrm{\AA}$) than the sdB star (9.37 $\mathrm{\AA}$) due to 
its higher  gravity. 

\begin{figure*}
\centering
\begin{minipage}[c]{0.5\textwidth}
\includegraphics[width=80mm]{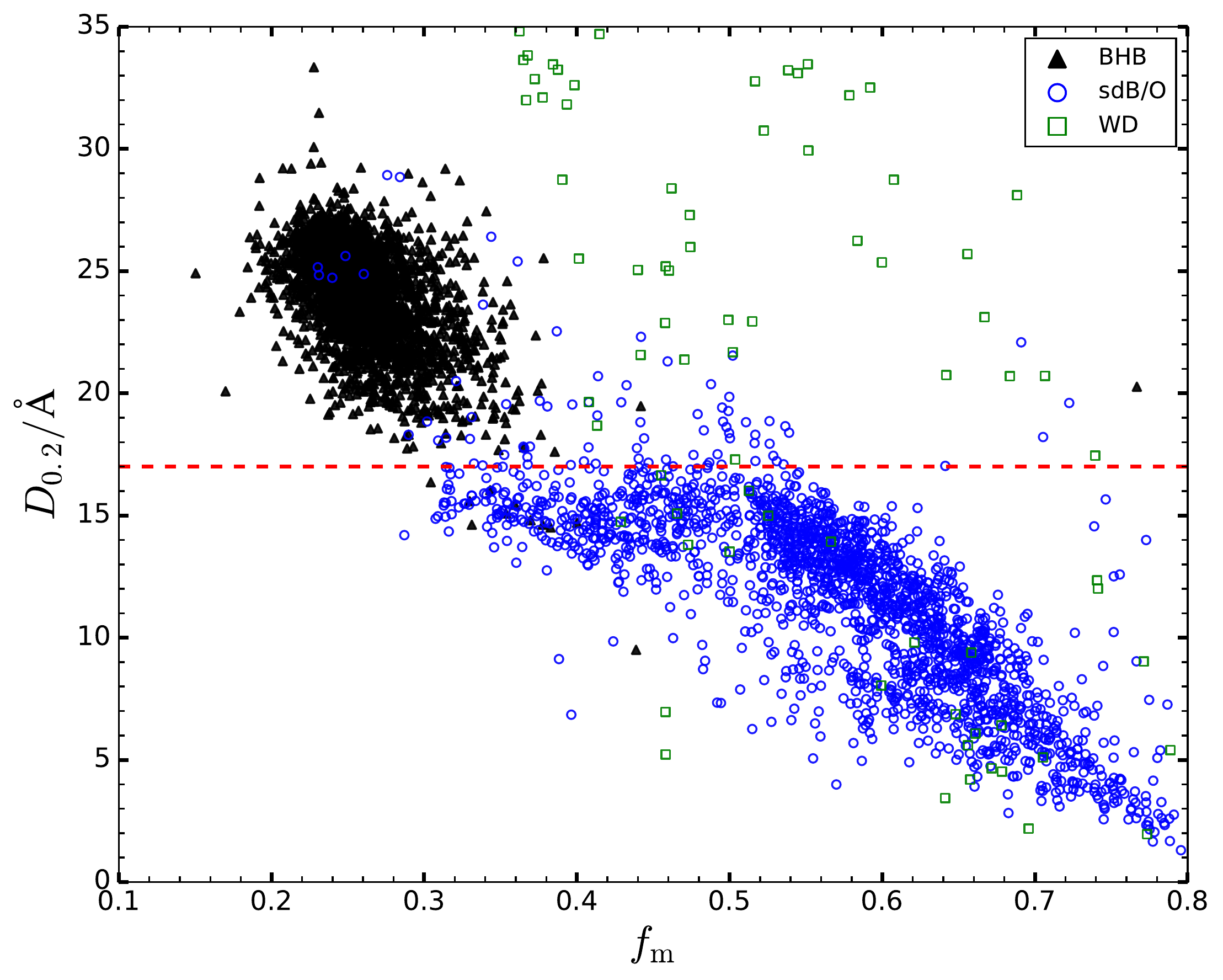}
\centerline{(a)}
\end{minipage}%
\centering
\begin{minipage}[c]{0.5\textwidth}
\includegraphics[width=80mm]{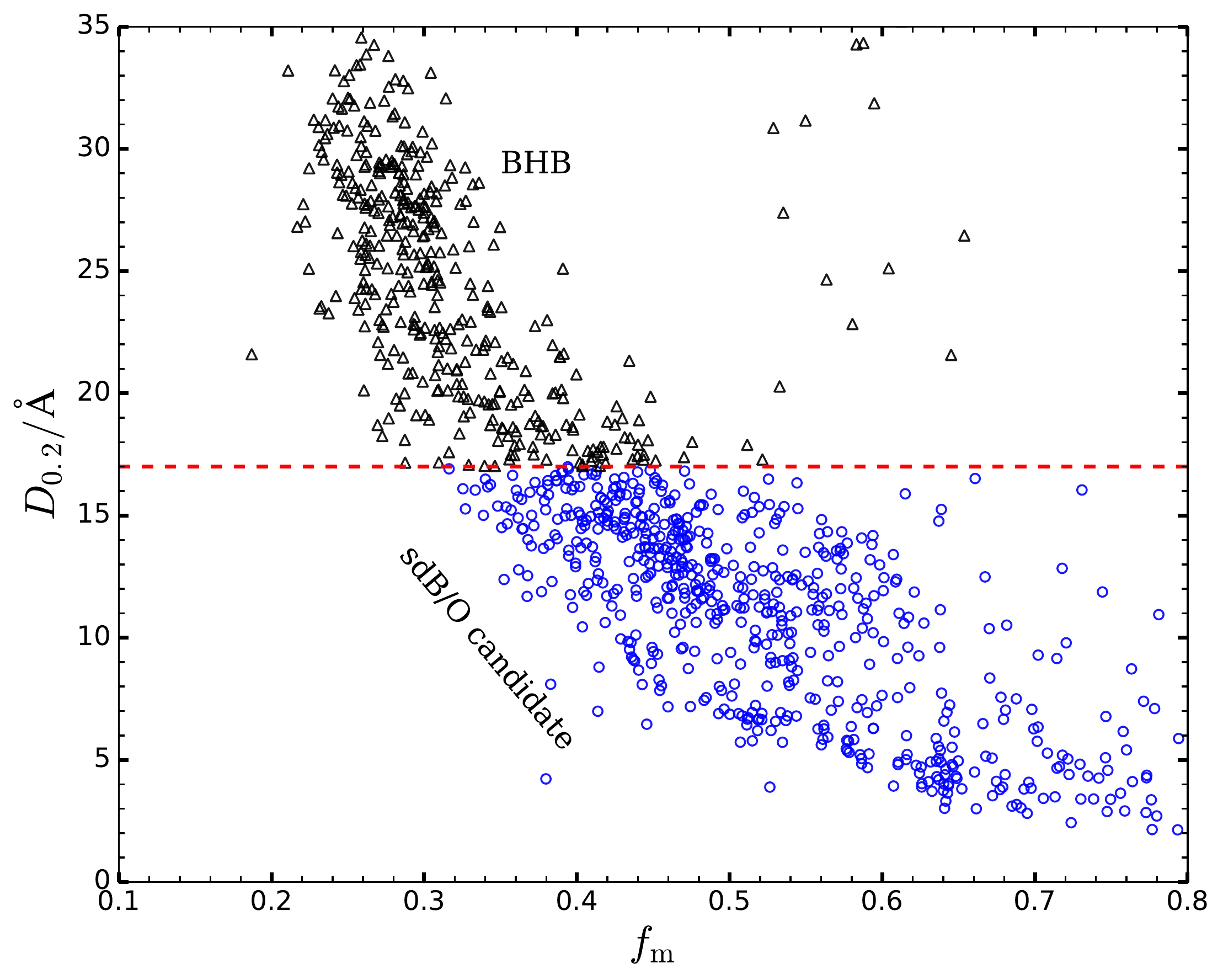}
\centerline{(b)}
\end{minipage}%
\caption{Panel (a): the distribution of BHB stars, hot subdwarfs  and WDs in the $D_{\rm 0.2}-f_{\rm m}$ diagram. 
Panel (b): Our hot subdwarf candidates selected by the HELM algorithm in the $D_{\rm 0.2}-f_{\rm m}$ diagram. The red dashed line is a clear boundary between BHB stars and hot subdwarfs at $D_{0.2}$\ =17.0 \AA. }
\end{figure*}

To better demonstrate the differences of  $D_{0.2}$ and $f_\mathrm{m}$ among BHB stars, hot 
subdwarfs and WDs, we selected some 
known BHB stars,  hot subdwarfs and WDs from 
published catalogues and put them into the 
$D_{0.2}$ - $f_\mathrm{m}$ diagram in Panel (a) of Fig 2. 
Black solid triangles denote BHB stars identified from 
Xue et al. (2008), blue open circles represent hot subdwarfs selected 
from the catalogue of Geier et al. (2017), and  
green open squares are WDs from Eisenstein et al. (2006). 
BHB stars are concentrated quite well in the upper left corner of Panel (a),  
and subdwarfs distribute in a strip from the middle center to the bottom right,  
while  WDs locate on the upper right and middle area of the panel 
(note that most of of the selected WDs have $D_{\rm 0.2}$ 
values larger than 35 \AA\ and are off the panel). 
As expected, there is a remarkable gap between BHB stars and 
hot subdwarf stars near $D_{0.2}$\ = 17.0 \AA\, which is marked by the  
red dashed horizontal line in Panel (a). Since WDs present much  
larger values of $D_{0.2}$ than BHB and hot subdwarf stars,  
$D_{0.2}$\ = 17.0  can be used as a criterion to  distinguish hot subdwarf 
stars from  BHB stars and WDs  in our sample.

Panel (b) of Fig 2 shows the values 
of $D_{0.2}$ and $f_\mathrm{m}$ for the 1034 sample spectra 
selected by HELM (see Section 2 and Paper I).  
To compare with Panel (a) clearly, we plot a dashed horizontal  
line at $D_{0.2}$ = 17.0 \AA\ in Panel (b) as well, which denotes the gap between BHB stars 
and hot subdwarf stars in Panel (a).  
Our sample in Panel (b) shows an analogous distribution to the stars in Panel (a), 
with the notable exception that the obvious gap at $D_{\rm 0.2}\ =\ 17.0$ \AA\ is not seen in Panel (b).  
This is due to the fact that the selected BHB stars in Panel (a) are stars with  temperatures  
in the range of $T_{\rm eff}\ =\ 7000 - 10\,000$ and surface gravity 
in a  range of $\log{g}\ =\ 2.5-4.0$ cm\,s$^{-2}$ (Xue et al. 2008), which are much lower than 
the temperatures and gravities of hot subdwarf stars 
(e.g., $T_\mathrm{eff}\geq$\ 20\,000$ K$ and log$\ g\geq 5.0 \ \mathrm{cm\,s}^{-2}$, Heber 2016), 
while the stars selected by HELM form a more evenly distributed mix of stars  
and the gap in the $D_{\rm 0.2}-f_{\rm m}$ diagram is filled up.  Therefore, our sample 
contains not only BHB stars with low temperatures, hot subdwarf stars and WDs, 
but also includes high temperature BHB stars (e.g., 10 000 - 20 000 K) and B-MS stars, because 
these stars present  similar temperatures to hot subdwarf stars in 
lower temperatures (e.g., He-poor sdB stars). 
Therefore, high temperature BHB stars  and B-MS stars 
fill the gap presented in Panel (a) and make a continuous distribution for  our sample 
in $D_{0.2}$ - $f_\mathrm{m}$ diagram. Note 
that there are a few  stars in the upper right and middle area of Panel (b), which are typically  
occupied by WDs in Panel (a). This demonstrates that 
a few WDs are in our sample, and HELM is  very efficient at
distinguishing hot subdwarf stars from WDs.  
 Nevertheless, the criterion  of $D_{0.2}$ = 17.0 \AA\ still 
 excludes most  BHB stars with low temperatures and WDs, while preserving hot subdwarf stars 
 in our sample. 
 
After applying the selection criterion of $D_{0.2}<17.0 $ \AA\,  
we obtained 578 hot subdwarf candidate spectra, among which 161 spectra present obvious Mg I triplet lines at 5170 $\mathrm{\AA}$ or Ca II triplet lines at 8650 $\mathrm{\AA}$. These lines are characteristic of
cool stars and such subdwarfs are double-lined composite spectrum binary candidates, that will be studied in a forthcoming publication.  
Therefore, our  hot subdwarf sample selected by $D_{0.2}$-$f_\mathrm{m}$ method consists of 417 spectra, for which the atmospheric parameters were determined by fitting their H Balmer and He lines.  

The $D_{0.2}$-$f_\mathrm{m}$  method is able to exclude most of the BHB stars and WDs in our sample. However, 
as the method is based on measuring the width and depth of H$_\delta$ line,  
some hot subdwarfs with weak or no obvious H$_\delta$ lines (e.g., He-sdO, He-sdB) could be also  removed 
from our sample. Furthermore, the values of $D_{0.2}$ and $f_\mathrm{m}$ for some spectra   
are  difficult to obtain from poor quality spectra near the  H$_\delta$ line.  
To assess the completeness of our sample we used {\sc XTgrid} (N\'emeth et al. 2012; Vennes et al. 2011, see next section for detail) to make a spectral classification for the 456 spectra which were removed by the $D_{0.2}$-$f_\mathrm{m}$ method. With this procedure we could recover  a further 48 hot subdwarf candidates from low quality spectra. The atmospheric parameters of these 48 spectra together with the 417 spectra selected by $D_{0.2}$-$f_\mathrm{m}$  method (i.e., 465 spectra in total) are determined by fitting their LAMOST optical spectra with synthetic spectra (see next section). All objects with atmospheric parameters  characteristic of hot subdwarfs were selected 
as hot subdwarf candidates.

 \subsection{Atmospheric parameters of hot subdwarf candidates }

\begin{figure}
\centering
\includegraphics [width=100mm]{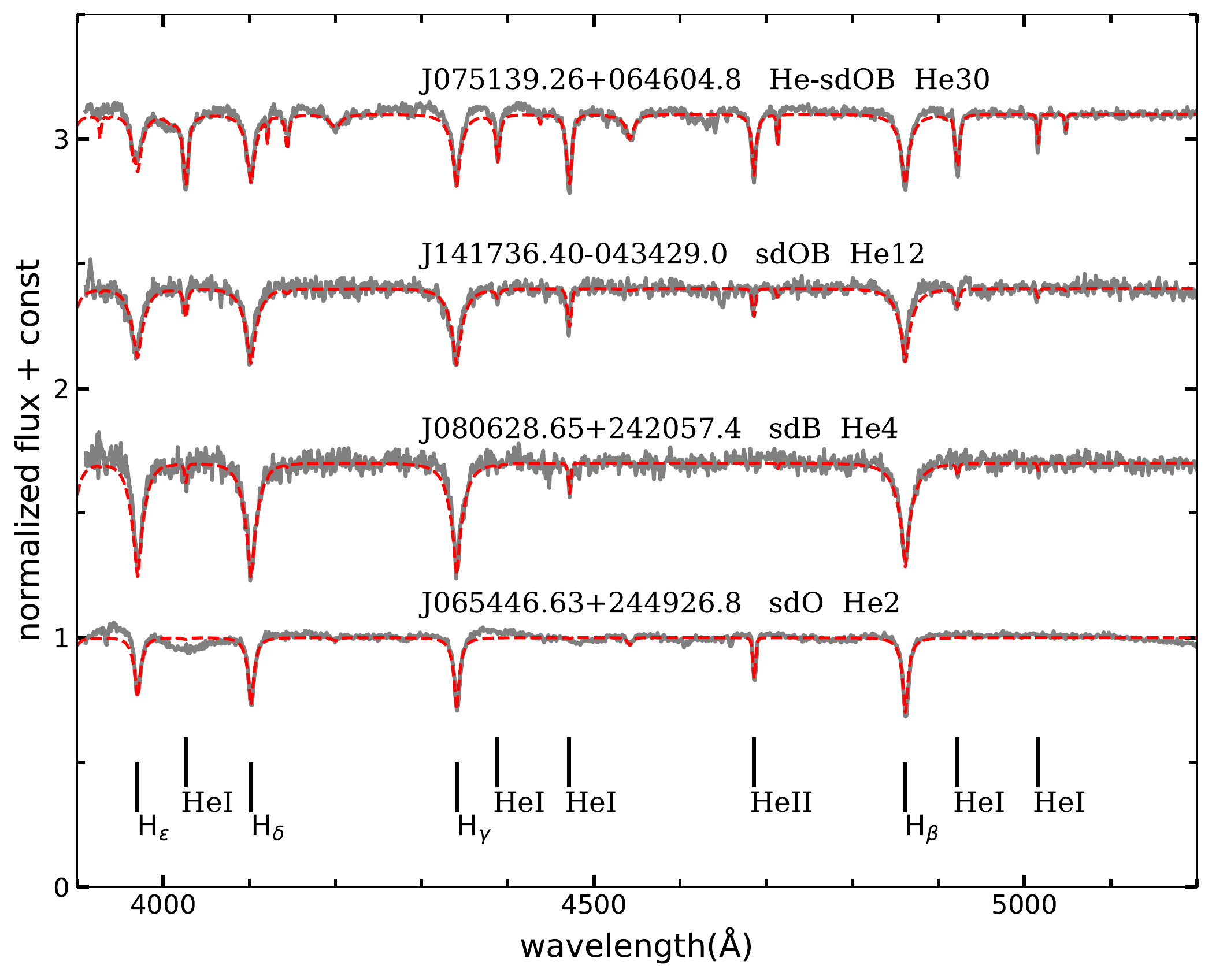}

\caption{Four normalized spectra of hot subdwarf stars with different spectral types identified in this study. Best-fitting synthetic spectra are over plotted by a red dashed line on each spectra.  From top  to bottom,  a He-sdOB, sdOB, sdB and sdO star is presented respectively. Some H Balmer  lines and important He I and 
He II lines marked at the bottom of the figure. }
\end{figure}

To determine the atmospheric parameters of the final hot subdwarf 
sample we fitted NLTE models to the observations. 
We used the NLTE model atmosphere code {\sc Tlusty} (version 204; Hubeny \& Lanz (2017) to calculate models with H and He composition and corresponding  synthetic spectra with {\sc Synspec} (version 49; Lanz et al. 2007). 
Details of the model calculations are described by  N\'emeth et al. (2014). 
The spectral analysis was done by a steepest-descent iterative  $\chi^2$ minimization procedure, 
which is implemented in the fitting program {\sc XTgrid} (N\'emeth et al. 2012; Vennes et al. 2011). 
This algorithm fits the entire optical range and attempts to reproduce the observed line profiles simultaneously. 
Final parameter errors are determined by departing from the best fitting parameters in one dimension until the  statistical limit for the 60\% confidence level of a single parameter is reached, 
separately for positive and negative  error bars. To match the resolution of LAMOST spectra we convolved the synthetic spectra with a Gaussian profile  at a constant resolution ($R\ =\ 1800$).  

Fig 3 shows the best fitting models for  four representative hot subdwarf spectra from our  
sample. In this figure, gray solid curves denote the normalized stellar 
spectra\footnote{The continuum for each spectrum was fitted automatically in {\sc XTgrid}}, 
while red dashed curves represent the best fitting synthetic spectra. The positions of  
the strongest  H Balmer lines, He I and He II lines  are  marked in Fig 3 as well. 
The label 'He' plus an integer for each spectrum is the helium class 
following the hot subdwarf classification scheme of Drilling et al. (2013), which is based on 
He line strength (see Sect 4 for details). 
The top spectrum is a He-sdOB star with dominant He I lines and weak 
H Balmer lines, while 
the second spectrum from the top is a  sdOB star, which shows 
dominant H Balmer lines with both weak He I and He II lines.  
The third spectrum from the top is a typical sdB star,  
which presents broad H Balmer lines with weak He I  lines. Finally, the spectrum  
at the bottom of the figure is classified  as a sdO star, because of its dominant H 
Balmer lines with weak He II line at 4686 $\mathrm{\AA}$ while no He l lines can be detected.

By employing {\sc XTgrid} , we obtained the atmospheric parameters 
(e.g., $T_\mathrm{eff}$, $\mathrm{log}\ g$ and He abundance) 
for the 465 spectra selected in Section 3. We classified stars with  $T_\mathrm{eff} \ge$ 20\,000 K 
and $\mathrm{log}\ g\ge$ 5.0  as hot subdwarf stars,  with 
$T_\mathrm{eff} <$ 20\,000K  and $\mathrm{log}g <$ 5.0 as hot BHB stars, while 
stars with $\mathrm{log}\ g <$ 4.5  as B-MS stars following the classification scheme of N\'emeth et al. (2012). 
After this procedure, we selected 76 hot subdwarf candidates based on 
their atmospheric parameters. We  checked our results by 
Gaia Hertzsprung-Russell diagram (HRD) in next section. 

\subsection{Cross matching our results with the HRD of Gaia DR2  }
\begin{figure}
\centering
\includegraphics [width=90mm]{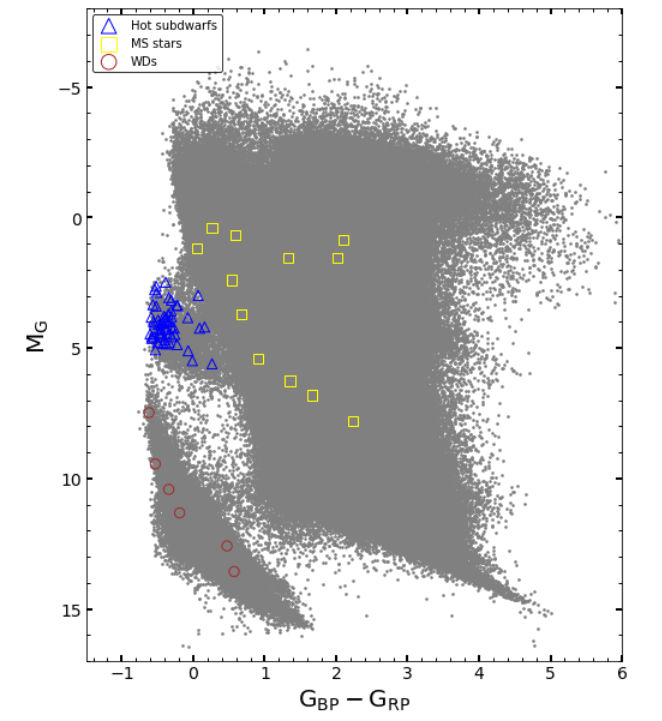}

\caption{The distribution of 74 selected subdwarf candidates in the HRD of Gaia DR2. 57 stars (marked with 
blue triangles) locate in the subdwarf region, and 12 stars (denoted by yellow squares) are distributed along the MS region, while the position of 6 stars (represented by red circles) correspond to the WD sequence.  }
\end{figure}

 The second data release (DR2) from Gaia (Gaia Collaboration et al. 2018a) 
provides high-precision astrometry and photometry for about 1.3 billion sources over 
the full sky. Based on this huge database, Gaia Collaboration et al. (2018b) 
built the Gaia DR2 HRD by using the 
most precise parallax and photometry (see Sect 2 in Gaia Collaboration et al. 2018b 
for their detailed selection filters). To check our final results, 
we cross matched the  76 hot subdwarf candidates 
with the database of Gaia DR2, and got 75 common objects within the radius of five arcseconds, 
among which one object had negative parallax, and it was removed from our sample. 
Fig 4 shows the HRD from Gaia Collaboration et al. (2018b) together with 
the 74 stars in common with this study. Gray dots denote the objects  from Gaia DR2 selected by  
Gaia Collaboration (65\,921\,112 stars in total, see Fig 1 of Gaia Collaboration et al. 2018b), 
while blue triangles, yellow squares and red circles are the  common stars in our sample. 
We found 56 stars (e.g., blue triangles) to be located in the hot subdwarf region of the HRD. 
Therefore, these 56 objects are finally identified as  
hot subdwarf stars in this study. On the other hand, we found 12 stars (e.g., yellow squares) distributed   
along the wide MS\footnote{Extinction is not considered in 
the HRD of Fig 1 in Gaia Collaboration et al. (2018b), therefore the MS is  
wider and can not be distinguished very clearly from the RGB. But the WD  and hot 
subdwarf sequences are presented more clearly in this HRD.} , and 6 stars (e.g., red circles) 
are along the WD sequence.

\section{Results}
Using the method described in Section 3, we identified 56 hot subdwarf stars. 
We followed the spectral classification scheme in 
Moehler et al. (1990) and Geier et al. (2017) to classify hot subdwarf stars: 
stars showing strong H Balmer lines with weak or no He I lines are classified as sdB stars; 
stars showing strong H Balmer lines accompanied by He II absorption are considered as 
sdO stars; stars having H Balmer lines accompanied both by weak He I and He II lines are 
classified as sdOB stars and stars with dominant He I lines and weak H Balmer lines are He-sdOB stars, while stars with 
dominant He II lines are He-sdO stars. Based on this simple classification scheme, we identified 
31 sdB stars, 11 sdO stars, 9 sdOB stars, 4 He-sdOB and 1 He-sdO stars.

 Drilling et al. (2013) designed  an MK (Morgan-Keenan)-like 
system of spectral classification for hot subdwarf stars, in which 
a spectral class, a luminosity class and a helium class are used to classify 
hot subdwarf stars. The spectral class is based on the 
MK standards of spectral classes O and B stars, and the 
luminosity class is based on the H and He line widths (see 
Sect 3 in Drilling et al. 2013). On the other hand, 
the helium class is described by an integer from 0 to 40 
denoting the strengths of the He lines relative to the H Balmer lines,  and it 
is roughly equal to the following function of the relative line depths: 
\begin{equation}
20\ \frac{\mathrm{HeI}\ \lambda4471+\mathrm{HeII}\ \lambda4541}
{\mathrm{H}_{\gamma}-0.83\ \mathrm{HeII}\ \lambda4541}
\end{equation}
for helium class 0-20, and 
\begin{equation}
40-20\ \frac{\mathrm{H}_{\gamma}-0.83\ \mathrm{HeII}\ \lambda4541}
{\mathrm{HeI}\ \lambda4471+\mathrm{HeII}\ \lambda4541}
\end{equation}
for helium class 20-40. We also appended this helium class  
for our  hot subdwarf stars (see Table 1).  

The atmospheric parameters of the 56 identified hot subdwarf stars  together with 
the information of 12 MS stars and 6  WDs are listed in Table 1.   
The atmospheric parameters of the MS stars and WDs are not presented. 
In column 1-11 of Table 1, we have presented the LAMOST designation, right ascension, 
declination, effective temperature , surface gravity and  He abundance obtained in 
this study, spectral classification type, SNR in the \textit{u} band, 
apparent magnitudes in the u and \textit{g} band of SDSS,  apparent 
magnitudes in the \textit{G} band of Gaia DR2, respectively. 
We also cross-matched our hot subdwarf stars with the hot subdwarfs list 
in Geier et al. (2017) and N\'emeth et al. (2012). In Table 1, 
the common hot subdwarf stars with Geier et al. (2017) are 
labeled by $^{*}$, and the common hot subdwarf stars with N\'emeth et al. (2012) are 
marked by $^{\dagger}$.

\begin{table*}
\scriptsize
%\centering
 \begin{minipage}{180mm}
 \caption{Information for the 74 stars analyzed in this study. From left to right of the table, it gives the LAMOST designation of the objects, right ascension, declination,  effective temperature, gravity,  helium abundance, spectral classification type, SNR in  \textit{u} band, apparent magnitude in \textit{u} and \textit{g} band from SDSS and apparent magnitude in \textit{G} band from Gaia DR2, respectively.  }
 \end{minipage}\\
\centering
    \begin{tabularx}{18.0cm}{lllcccccccccccX}
\hline\noalign{\smallskip}
 Designation$^ a$  & ra$^ b$   & dec  & $T_\mathrm{eff}$ &  $\mathrm{log}\ g$ & $\mathrm{log}(n\mathrm{He}/n\mathrm{H})^c$ &Sptype & SNR &uSDSS &gSDSS  &G\ GaiaDR2 & \\
 LAMOST & LAMOST &  LAMOST&$(K)$&($\mathrm{cm\ s^{-2}}$)& & &\textit{u}-band &(mag) &(mag) &(mag)  &\\
\hline\noalign{\smallskip}
J002124.79+402857.1 & 5.3532989 & 40.482537 & 25850$\pm$\ 580 & 5.42$\pm$0.11 & -2.57$\pm$0.18 & sdB He4 & 18.7 & \ \ - & 15.19 & 15.51 & \\ 
J002355.23+420905.5$^{*}$ & 5.9801396 & 42.151544 & 30150$\pm$\ 280 & 5.47$\pm$0.06 & -2.31$\pm$0.14 & sdB He6 & 18.6 & \ \ - & 15.46 & 15.79 & \\ 
J003627.19+271000.7 & 9.113308 & 27.166863 & \ \ - & \ \ - & \ \ - & MS  & 45.0 & 14.93 & 14.67 & 14.64 & \\ 
J003801.72+343156.2 & 9.5071771 & 34.53228 & 40850$\pm$\ 610 & 5.49$\pm$0.10 & -0.23$\pm$0.09 & He-sdOB He33 & 17.9 & \ \ - & 13.66 & 13.90 & \\ 
J004949.26+352200.9$^{*}$ & 12.455266 & 35.366938 & 34960$\pm$\ 690 & 5.83$\pm$0.12 & -1.49$\pm$0.10 & sdOB He13 & 25.3 & \ \ - & 14.54 & 14.82 & \\ 
J010448.81+362742.4$^{*}$ & 16.203409 & 36.461784 & 32260$\pm$\ 60 & 5.74$\pm$0.02 & -1.63$\pm$0.03 & sdOB He11 & 90.6 & 12.55 & 12.95 & 12.40 & \\ 
J010945.73+374538.5$^{*}$ & 17.440552 & 37.760704 & 29980$\pm$\ 100 & 5.49$\pm$0.03 & -3.54$\pm$0.26 & sdB He2 & 25.6 & 13.96 & 14.61 & 13.87 & \\ 
J011857.19-002545.5$^{*}$ & 19.738333 & -0.429333 & 29060$\pm$\ 140 & 5.48$\pm$0.04 & -3.16$\pm$0.25 & sdB He2 & 15.7 & 14.49 & 14.60 & 14.82 & \\ 
J013134.51+323723.7 & 22.893792 & 32.623252 & 60390$\pm$\ 720 & 5.48$\pm$0.05 & -1.40$\pm$0.10 & sdO He8 & 10.9 & \ \ - & 15.00 & 15.30 & \\ 
J014710.62+303213.2 & 26.794254 & 30.537002 & 22110$\pm$\ 210 & 5.00$\pm$0.07 & -2.05$\pm$0.12 & sdB He6 & 18.8 & \ \ - & 14.10 & 14.35 & \\ 
J015054.28+310746.7 & 27.72618 & 31.129651 & 28540$\pm$\ 180 & 5.70$\pm$0.04 & -1.69$\pm$0.05 & sdB He10 & 16.9 & \ \ - & 13.97 & 14.32 & \\ 
J020932.45+430712.5$^{*}$ & 32.385219 & 43.12014 & 27580$\pm$\ 500 & 5.42$\pm$0.03 & -2.73$\pm$0.16 & sdB He5 & 11.8 & 14.42 & 14.86 & 14.34 & \\ 
J022517.07+031218.2 & 36.3211422 & 3.2050785 & \ \ - & \ \ - & \ \ - & WD  & 15.1 & 16.24 & 16.70 & 16.95 & \\ 
J023551.35+011845.1 & 38.963972 & 1.312544 & \ \ - & \ \ - & \ \ - & WD  & 10.4 & 16.97 & 16.41 & 16.17 & \\ 
J030025.22+003224.3 & 45.10512 & 0.54009 & \ \ - & \ \ - & \ \ - & MS  & 13.1 & 23.89 & 21.76 & 20.36 & \\ 
J031756.92+322950.4 & 49.487181 & 32.497341 & 33860$\pm$\ 430 & 6.07$\pm$0.15 & -1.62$\pm$0.12 & sdB He13 & 15.9 & \ \ - & 15.58 & 15.72 & \\ 
J035926.96+270508.6 & 59.862336 & 27.08573 & 35160$\pm$\ 380 & 5.51$\pm$0.04 & -2.74$\pm$0.35 & sdOB He2 & 14.0 & \ \ - & 14.97 & 15.10 & \\ 
J040613.24+465133.6 & 61.555205 & 46.859349 & \ \ - & \ \ - & \ \ - & MS  & 15.2 & 14.77 & \ \ - & 14.59 & \\ 
J051425.36+332344.3 & 78.605685 & 33.395662 & \ \ - & \ \ - & \ \ - & MS  & 10.4 & \ \ - & 15.04 & 13.38 & \\ 
J053656.48+395518.7$^{*}$ & 84.235335 & 39.92188 & 38490$\pm$\ 350 & 5.54$\pm$0.07 & -0.65$\pm$0.07 & sdOB He16 & 14.7 & \ \ - & 13.67 & 13.92 & \\ 
J054447.48+272032.0 & 86.197835 & 27.342228 & \ \ - & \ \ - & \ \ - & WD  & 10.3 & \ \ - & 17.08 & 16.93 & \\ 
J055151.32+220437.0 & 87.96384 & 22.076954 & 29610$\pm$\ 110 & 5.66$\pm$0.03 & -2.22$\pm$0.05 & sdB He5 & 24.4 & \ \ - & 12.85 & 13.17 & \\ 
J055227.67+155311.4 & 88.115311 & 15.886516 & \ \ - & \ \ - & \ \ - & WD  & 23.1 & \ \ - & 12.52 & 13.03 & \\ 
J055348.85+325601.7 & 88.453581 & 32.93382 & 30490$\pm$\ 110 & 5.68$\pm$0.02 & -2.15$\pm$0.04 & sdB He5 & 44.0 & \ \ - & 14.02 & 14.17 & \\ 
J055411.88+220459.7 & 88.549534 & 22.083273 & \ \ - & \ \ - & \ \ - & MS  & 10.2 & \ \ - & 13.28 & 13.17 & \\ 
J055926.92+271321.0 & 89.862203 & 27.222502 & \ \ - & \ \ - & \ \ - & MS  & 10.9 & \ \ - & 19.20 & 17.99 & \\ 
J062704.91+345809.5$^{*}$ & 96.770481 & 34.969325 & 25080$\pm$\ 380 & 5.26$\pm$0.08 & -3.57$\pm$0.62 & sdB He1 & 10.8 & \ \ - & 14.19 & 14.43 & \\ 
J062836.51+325031.5 & 97.152155 & 32.842084 & 42740$\pm$\ 570 & 5.30$\pm$0.12 & 0.20$\pm$0.10 & He-sdOB He37 & 21.5 & \ \ - & 14.51 & 14.71 & \\ 
J063210.36+281041.7 & 98.043207 & 28.178276 & 45130$\pm$\ 330 & 5.51$\pm$0.12 & 0.33$\pm$0.06 & He-sdOB He40 & 17.7 & \ \ - & 14.82 & 15.10 & \\ 
J063526.61+323109.8 & 98.86089 & 32.519401 & \ \ - & \ \ - & \ \ - & MS  & 11.6 & \ \ - & 15.95 & 15.15 & \\ 
J063952.15+515700.9 & 99.967315 & 51.950267 & 29720$\pm$\ 110 & 5.37$\pm$0.04 & -3.00$\pm$0.73 & sdB He1 & 35.8 & \ \ - & \ \ - & 11.96 & \\ 
J064618.36+292013.2$^{*}$ & 101.57652 & 29.337016 & 38740$\pm$\ 450 & 5.90$\pm$0.05 & $-4.00>$ & sdO He0 & 73.4 & \ \ - & \ \ - & 13.59 & \\ 
J064814.13+171056.2 & 102.05891 & 17.182305 & \ \ - & \ \ - & \ \ - & MS  & 10.6 & \ \ - & 14.96 & 13.23 & \\ 
J065446.63+244926.8 & 103.69431 & 24.82412 & 58700$\pm$3600 & 5.17$\pm$0.05 & -2.04$\pm$0.10 & sdO He2 & 55.7 & \ \ - & 13.65 & 13.99 & \\ 
J065532.98+220349.6 & 103.88743 & 22.063784 & 45090$\pm$\ 890 & 5.62$\pm$0.05 & -1.71$\pm$0.08 & sdO He6 & 30.5 & \ \ - & \ \ - & 13.70 & \\ 
J065647.77+242958.8 & 104.19908 & 24.499685 & \ \ - & \ \ - & \ \ - & MS  & 18.7 & \ \ - & \ \ - & 10.19 & \\ 
J065748.42+253251.1 & 104.45177 & 25.547541 & 44930$\pm$1160 & 6.48$\pm$0.10 & $-4.00>$ & sdB He19 & 16.1 & \ \ - & 15.89 & 16.05 & \\ 
J065816.71+094343.1 & 104.56965 & 9.7286415 & 36270$\pm$\ 320 & 5.03$\pm$0.03 & -1.70$\pm$0.08 & sdOB He11 & 17.1 & \ \ - & 13.27 & 13.59 & \\ 
J070619.19+242910.5 & 106.57996 & 24.486267 & 61820$\pm$6030 & 5.30$\pm$0.04 & -2.00$\pm$0.13 & sdO He4 & 15.0 & \ \ - & 15.77 & 15.81 & \\ 
J071202.40+113332.4 & 108.01 & 11.559014 & 24720$\pm$\ 180 & 5.10$\pm$0.04 & -2.63$\pm$0.07 & sdB He5 & 33.0 & \ \ - & \ \ - & 12.46 & \\ 
\hline\noalign{\smallskip} 
\end{tabularx}\\
{$^a$ Stars labeled with $\ast$ also appear in the hot subdwarf catalogue of Geier et al. (2017).}\\
{$^b$ Stars labeled with $\dagger$ also appear in N\'emeth et al. (2012).} \\
{$^c$ "$>$" denotes a upper limit of $\mathrm{log}(n\mathrm{He}/n\mathrm{H})$ for the object.}\\
\end{table*}

\setcounter{table}{0}
\begin{table*}
\scriptsize
%\centering
 \begin{minipage}{180mm}
 \caption{Continued }
 \end{minipage}\\
\centering
    \begin{tabularx}{18.0cm}{lllcccccccccccX}
\hline\noalign{\smallskip}
 Designation$^ a$  & ra$^ b$   & dec  & $T_\mathrm{eff}$ &  $\mathrm{log}\ g$ & $\mathrm{log}(n\mathrm{He}/n\mathrm{H})^c$ &Sptype & SNR &uSDSS &gSDSS  &G\ GaiaDR2 & \\
 LAMOST & LAMOST &  LAMOST&$(K)$&($\mathrm{cm\ s^{-2}}$)& & &\textit{u}-band &(mag) &(mag) &(mag)  &\\
\hline\noalign{\smallskip}
J072835.11+280239.1 & 112.1463 & 28.044199 & 86250$\pm$16170 & 5.77$\pm$0.16 & 0.04$\pm$0.12 & He-sdO He40 & 10.1 & \ \ - & 15.45 & 15.78 & \\ 
J073446.14+342120.8 & 113.69226 & 34.355805 & 25510$\pm$\ 680 & 5.15$\pm$0.07 & -2.42$\pm$0.09 & sdB He6 & 20.6 & \ \ - & 15.20 & 15.46 & \\ 
J073756.25+311646.5 & 114.48439 & 31.279597 & 30600$\pm$\ 130 & 5.45$\pm$0.03 & -2.47$\pm$0.12 & sdB He5 & 11.2 & \ \ - & \ \ - & 13.58 & \\ 
J074121.90+265425.8$^{*}$ & 115.34127 & 26.907168 & 29530$\pm$\ 460 & 5.30$\pm$0.07 & $-4.00>$ & sdB  & 11.6 & \ \ - & 15.52 & 15.59 & \\ 
J074435.14+302108.7$^{*}$ & 116.14643$^{\dagger}$ & 30.352421 & 28980$\pm$\ 200 & 5.51$\pm$0.03 & -2.95$\pm$0.10 & sdB He3 & 30.6 & \ \ - & 14.39 & 14.74 & \\ 
J074855.82+304247.0$^{*}$ & 117.23262$^{\dagger}$ & 30.713059 & 30910$\pm$\ 110 & 5.80$\pm$0.03 & -2.02$\pm$0.04 & sdB He4 & 31.8 & \ \ - & 13.76 & 14.06 & \\ 
J075139.26+064604.8 & 117.91362 & 6.7680011 & 39850$\pm$\ 180 & 5.61$\pm$0.04 & -0.16$\pm$0.03 & He-sdOB He30 & 39.9 & \ \ - & 13.21 & 13.50 & \\ 
J075412.37+294957.0$^{*}$ & 118.55157 & 29.832504 & 30910$\pm$1230 & 5.77$\pm$0.28 & -1.87$\pm$0.32 & sdB He7 & 14.5 & \ \ - & 14.24 & 14.57 & \\ 
J075922.99+164601.6 & 119.845827 & 16.767125 & 37930$\pm$\ 920 & 5.25$\pm$0.05 & -2.89$\pm$0.27 & sdO He1 & 23.9 & 13.84 & 14.94 & 14.42 & \\ 
J080327.92+342140.6$^{*}$ & 120.86637 & 34.361297 & 38130$\pm$1350 & 5.58$\pm$0.11 & -3.28$\pm$0.60 & sdO He3 & 26.1 & \ \ - & 14.75 & 15.06 & \\ 
J080611.66+334425.6 & 121.5486 & 33.740449 & \ \ - & \ \ - & \ \ - & WD  & 10.5 & \ \ - & 16.13 & 16.33 & \\ 
J080628.65+242057.4$^{*}$ & 121.61938 & 24.349293 & 27990$\pm$\ 350 & 5.48$\pm$0.04 & -2.50$\pm$0.14 & sdB He4 & 14.4 & \ \ - & 14.70 & 15.00 & \\ 
J080758.25+272434.3 & 121.99274 & 27.409538 & 38370$\pm$1190 & 5.58$\pm$0.08 & -3.41$\pm$0.66 & sdO He2 & 50.4 & \ \ - & 13.76 & 14.11 & \\ 
J084535.66+194150.2 & 131.3986$^{\dagger}$ & 19.697288 & 22070$\pm$\ 420 & 5.00$\pm$0.06 & -1.80$\pm$0.06 & sdB He7 & 18.4 & 13.13 & 13.49 & 13.26 & \\ 
J085649.36+170116.0$^{*}$ & 134.2056708$^{\dagger}$ & 17.021125 & 28810$\pm$\ 150 & 5.65$\pm$0.01 & -3.19$\pm$0.17 & sdB He2 & 56.3 & 14.67 & 12.73 & 12.81 & \\ 
J085851.11+021012.9$^{*}$ & 134.71299 & 2.1702667 & 48580$\pm$1150 & 5.61$\pm$0.07 & -1.83$\pm$0.09 & sdO He6 & 16.8 & \ \ - & 13.30 & 13.63 & \\ 
J093512.20+310959.3$^{*}$ & 143.8008625 & 31.166475 & 33870$\pm$\ 110 & 5.62$\pm$0.04 & -1.47$\pm$0.07 & sdOB He11 & 13.4 & 15.06 & 15.34 & 15.63 & \\ 
J112350.68+233645.8$^{*}$ & 170.961175 & 23.6127333 & 27560$\pm$\ 350 & 5.32$\pm$0.04 & -2.39$\pm$0.11 & sdB He5 & 15.8 & 13.76 & 13.90 & 14.15 & \\ 
J120624.36+570935.7$^{*}$ & 181.6015083$^{\dagger}$ & 57.1599222 & 34960$\pm$\ 230 & 5.70$\pm$0.04 & -1.81$\pm$0.06 & sdOB He9 & 18.4 & 14.28 & 14.60 & 14.85 & \\ 
J123652.66+501513.8$^{*}$ & 189.219429 & 50.253856 & 43250$\pm$2210 & 5.40$\pm$0.12 & -2.42$\pm$0.30 & sdO He2 & 22.7 & 13.96 & 14.38 & 14.65 & \\ 
J125229.60-030129.6$^{*}$ & 193.12335 & -3.0248924 & 30790$\pm$\ 480 & 5.59$\pm$0.09 & $-3.36>$ & sdB He0 & 13.4 & 15.46 & 15.71 & 15.65 & \\ 
J133640.95+515449.4 & 204.170631 & 51.913729 & 88450$\pm$21230 & 5.13$\pm$1.00 & -2.77$\pm$1.04 & sdOB - & 53.5 & 12.79 & 12.76 & 12.97 & \\ 
J135153.11-012946.6 & 207.9713167 & -1.4962778 & 31040$\pm$\ 560 & 6.03$\pm$0.12 & $-2.77>$ & sdB He0 & 11.2 & 15.31 & 15.45 & 15.66 & \\ 
J141736.40-043429.0 & 214.401671 & -4.574742 & 37750$\pm$\ 380 & 5.82$\pm$0.06 & -1.53$\pm$0.05 & sdOB He12 & 24.6 & 13.52 & 13.96 & 13.71 & \\ 
J144052.82-030852.6$^{*}$ & 220.220106 & -3.147965 & 29320$\pm$\ 40 & 5.44$\pm$0.03 & -2.74$\pm$0.05 & sdB He0 & 45.2 & 13.60 & 14.02 & 13.82 & \\ 
J161200.65+514943.5$^{*}$ & 243.0027458 & 51.82875 & 45130$\pm$1610 & 5.09$\pm$0.13 & -3.31$\pm$0.29 & sdB He2 & 10.9 & 13.26 & 13.54 & 13.67 & \\ 
J164718.35+322832.9 & 251.826491 & 32.475813 & \ \ - & \ \ - & \ \ - & WD  & 38.9 & 13.47 & 13.83 & 13.59 & \\ 
J171013.21+532646.0 & 257.555047 & 53.446121 & 28120$\pm$\ 340 & 5.83$\pm$0.03 & -2.42$\pm$0.12 & sdB He3 & 13.1 & 12.28 & 12.87 & 12.60 & \\ 
J171718.79+422609.2 & 259.32832 & 42.435913 & 55490$\pm$2130 & 5.78$\pm$0.03 & -3.01$\pm$0.29 & sdO He0 & 30.4 & 12.26 & 12.77 & 12.48 & \\ 
J175311.46+062541.5 & 268.2977592 & 6.4282084 & \ \ - & \ \ - & \ \ - & MS  & 11.6 & 14.68 & 13.66 & 14.58 & \\ 
J192216.18+405757.4 & 290.567417 & 40.965972 & \ \ - & \ \ - & \ \ - & MS  & 20.7 & 13.54 & \ \ - & 13.51 & \\ 
J192609.46+372008.1$^{*}$ & 291.539417 & 37.335611 & 31060$\pm$\ 240 & 5.97$\pm$0.04 & -1.65$\pm$0.04 & sdB He11 & 23.8 & 13.45 & \ \ - & 13.61 & \\ 
J213406.74+033415.4 & 323.528123 & 3.570953 & 40310$\pm$1390 & 6.12$\pm$0.12 & -1.60$\pm$0.18 & sdB - & 21.2 & 11.50 & 11.78 & 11.55 & \\ 
J223419.15+091620.5 & 338.57981 & 9.272378 & \ \ - & \ \ - & \ \ - & MS  & 13.2 & 13.89 & 13.93 & 13.87 & \\  
\hline\noalign{\smallskip} 
\end{tabularx}\\
\end{table*}

\subsection{Comparison with other studies}
 Among the 56 hot subdwarf stars in our study, 
25 stars have been already catalouged by Geier et al. (2017), and 5 stars are 
listed in N\'emeth et al. (2012). 
To check the results presented in our study, we compared the atmospheric 
parameters obtained in this study  with 
the ones from Geier et al. (2017) and N\'emeth et al. (2012) where their 
parameters are available.

We have 25 common hot subdwarf stars with Geier et al. (2017), but only 
11 stars with their $T_\mathrm{eff}$ and $\mathrm{log}\ g$  
are  available in the catalogue,   and 10  stars with their He 
abundances are available in the catalogue.  
The subplots from  left to right of Panel (a) in Fig 5 present the comparison of 
$T_\mathrm{eff}$, $\mathrm{log}\ g$ and  $\mathrm{log}(n\mathrm{He}/n\mathrm{H})$, 
respectively. As we see that both $T_\mathrm{eff}$ 
and $\mathrm{log}(n\mathrm{He}/n\mathrm{H})$ obtained in this 
study  matched well with the values from Geier et al. (2017). 
Although, the comparison of $\mathrm{log}\ g$ in the middle subplot of Panel (a) 
presents a more dispersive distribution than the other two parameters, but 
our results are still comparable with the values from literature. 

We also have 5 common hot subdwarf 
stars with N\'emeth et al. (2012), which are marked 
in Table 1. These stars  
are from the GALEX survey with low-resolution spectra. 
Similar as we see in Panel (a), both $T_\mathrm{eff}$ 
and $\mathrm{log}(n\mathrm{He}/n\mathrm{H})$ from 
this study match very well with the values from N\'emeth 
et al. (2012, see the left  and right subplots in Panel (b)). 
However, most of  the $\mathrm{log}\ g$ obtained in this study 
seem to be a little larger than the values from N\'emeth 
et al. (2012, see the middle subplot in Panel(b)). This could be due to the fact that the synthetic 
spectra used to fit the observed spectra in our study are 
calculated from atmospheric models only with H and He 
composition (N\'emeth et al. 2014), while the synthetic 
spectra used in N\'emeth et al. (2012) are calculated from atmospheric  
models not only with H and He composition but also 
include C, N and O composition. Furthermore, the observed 
spectra in our sample (obtained  in LAMOST survey) are 
different from the spectra in N\'emeth et al. (2012, obtained in 
GALEX survey), and the qualities (e.g., SNR) for the spectra are 
also different. Beyond these effects the major reason for the differences 
in the surface gravity is the inclusion of H Stark broadening tables 
from Tremblay \& Bergeron (2009)  
directly in the model atmosphere calculation in {\sc Tlusty} 
version 204, unlike in version 200 used by N\'emeth et al (2012).

\begin{figure*}
\centering
\begin{minipage}[c]{0.8\textwidth}
\includegraphics[width=140mm]{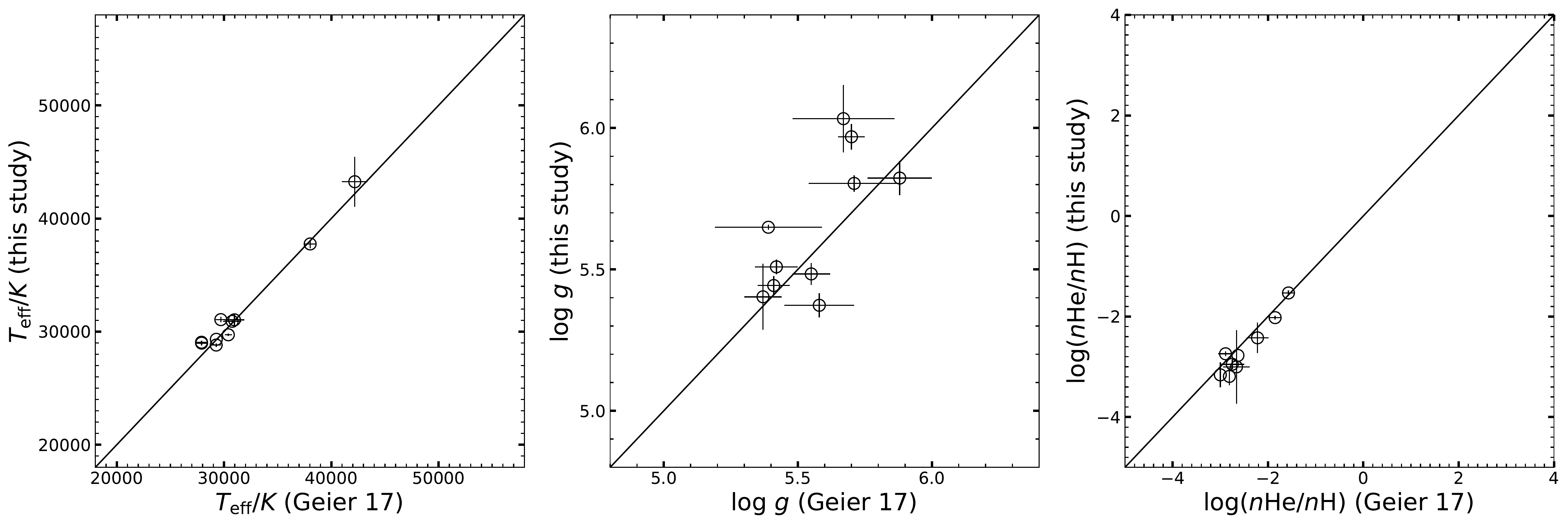}
\centerline{(a)}
\end{minipage}\\
\centering
\begin{minipage}[c]{0.8\textwidth}
\includegraphics[width=140mm]{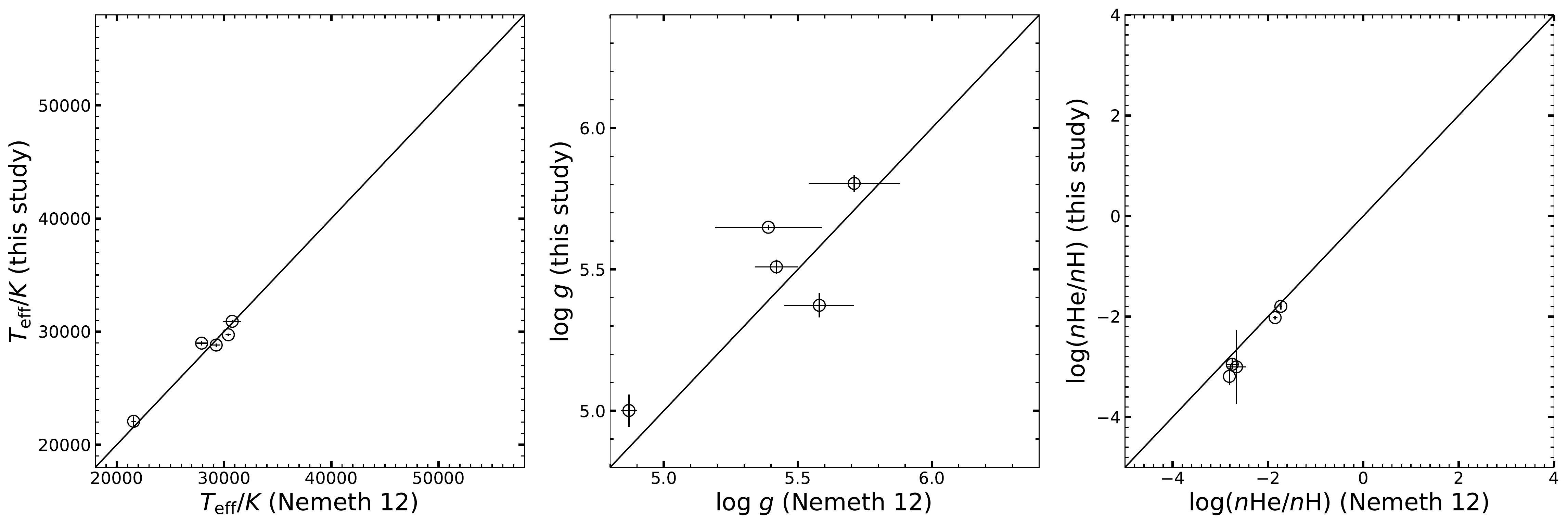}
\centerline{(b)}
\end{minipage}
\caption{Panel (a): Comparisons between  the atmospheric parameters    
obtained in this study and the ones from Geier et al. (2017). 
Panel (b): Comparisons between  the atmospheric parameters    
obtained in this study and the ones from N\'emeth et al. (2012).  }
\end{figure*}

\subsection{Parameter diagrams}
Fig 6 shows the distribution of all hot subdwarf stars from our study in the $T_{\rm eff}-\log\ {g}$ diagram. 
The thick solid line denotes the 
He main-sequence (He MS)  from Paczy\'nski (1971), while  
the two dashed lines represent the zero-age HB (ZAHB) and terminal-age HB (TAHB) 
for hot subdwarf stars with 
[Fe/H] = -1.48 from Dorman et al. (1993). The thin solid, dot-dashed and 
dotted curves  are the sdB evolution tracks from Han et al. (2002). From right to left, 
these sdB evolution tracks have the masses of 0.5, 0.6 and 0.7\ $\mathrm{M}_{\odot}$ respectively. 
The thin solid curves present a H-rich envelope mass of 0.0\ $\mathrm{M}_{\odot}$, 
the dot-dashed curves for  0.001\ $\mathrm{M}_{\odot}$, and the dotted curves for 0.005\ $\mathrm{M}_{\odot}$.  

\begin{figure}
\centering
\includegraphics [width=100mm]{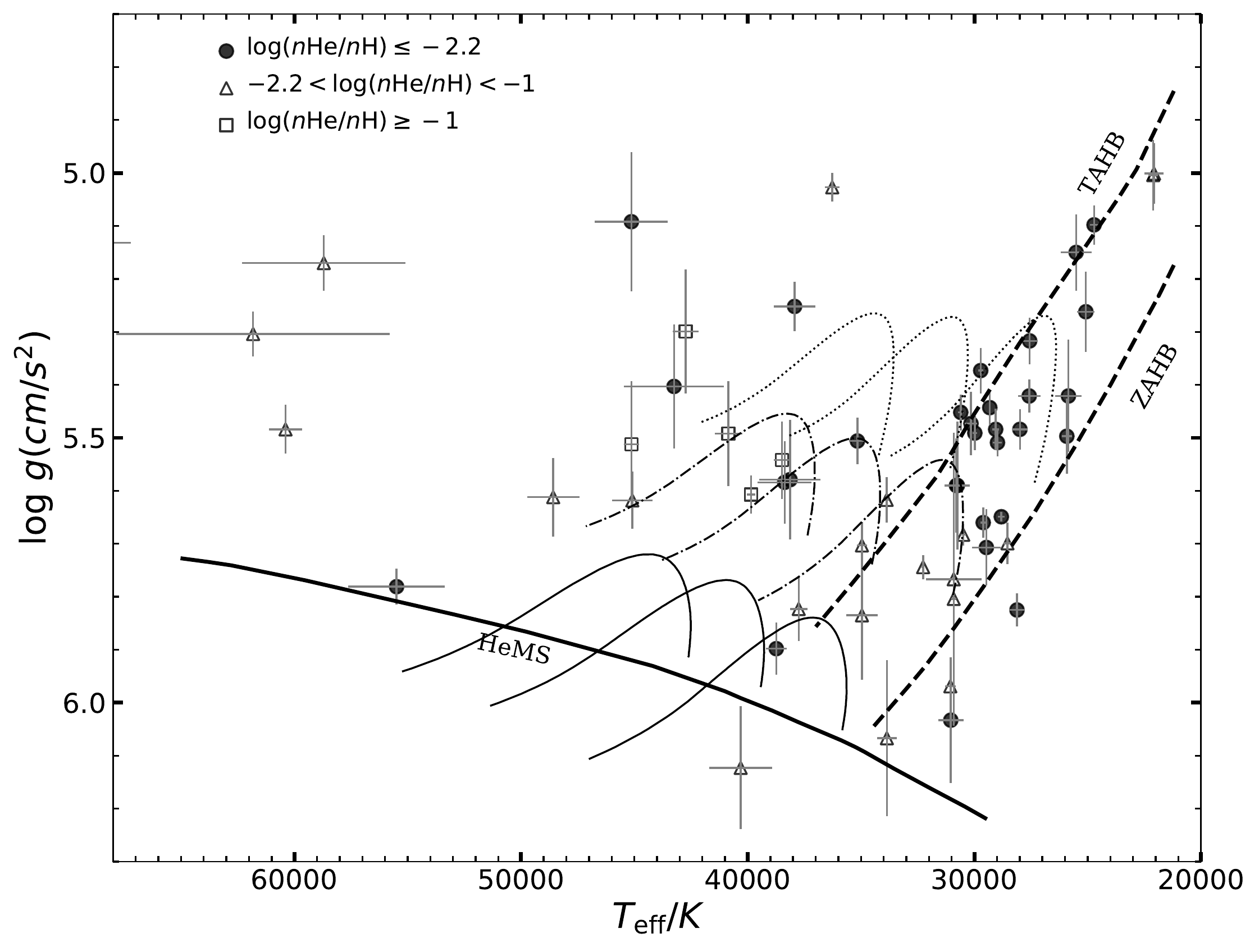}

\caption{$T_\mathrm{eff}$-$\mathrm{log}\ g$ diagram for  for the 56 hot subdwarf stars 
identified in this study. Stars with  $\mathrm{log}(n\mathrm{He}/n\mathrm{H})\leq-2.2$  are 
marked with filled circles, stars with  $-2.2< \mathrm{log}(n\mathrm{He}/n\mathrm{H})<-1.0$ are 
represented by open triangles, while stars with $\mathrm{log}(n\mathrm{He}/n\mathrm{H})\geq -1.0$
 are showed by open squares. The thick solid line denotes the He-MS from Paczy\'nski (1971), 
 and the two dashed lines represent ZAHB and TAHB from Dorman et al. (1993) with [Fe/H] = -1.48. 
 While the thin solid, dot-dashed, and dotted curves 
 represent the evolution tracks of hot subdwarf stars from Han et al. (2002).  See text for 
 the details on these evolution tracks.}
\end{figure}

We split our sample into three groups based on their He abundance following 
the scheme of N\'emeth et al. (2012). 
In Fig 6, filled circles denote hot subdwarf stars with 
their $\mathrm{log}(n\mathrm{He}/n\mathrm{H})\leq-2.2$. Most of 
these stars are He-poor sdB stars, and they are located near 
 $T_\mathrm{eff}$ = 29 000 K, and $\mathrm{log}\ g$ = 5.5 cm\,s$^{-2}$.   
A few of the stars in this He abundance range 
show very high temperatures (e.g., $T_\mathrm{eff}\geq$50 000 K), 
which suggests that they have already finished their core helium burning stage 
and now evolve towards the WD cooling tracks. 
Open triangles in Fig 6 represent hot subdwarf stars with 
$-2.2< \mathrm{log}(n\mathrm{He}/n\mathrm{H})<-1.0$. 
Most of these stars are found near $T_\mathrm{eff}$ = 32 000 K, 
and $\mathrm{log}g$ = 5.75 cm\,s$^{-2}$. 
These stars show higher gravities than previous 
group and  their temperatures show a  large dispersion. The third group contains stars 
with He abundances in the range of $-1.0\leq\mathrm{log}(n\mathrm{He}/n\mathrm{H})$, 
which are denoted by open squares in Fig 6. 
Actually, we just found five hot subdwarf stars in this He abundance range,  four of them are classified 
as He-sdOB stars and one is classified as He-sdO star based on our classification scheme. 

\begin{figure}
\centering
\includegraphics [width=100mm]{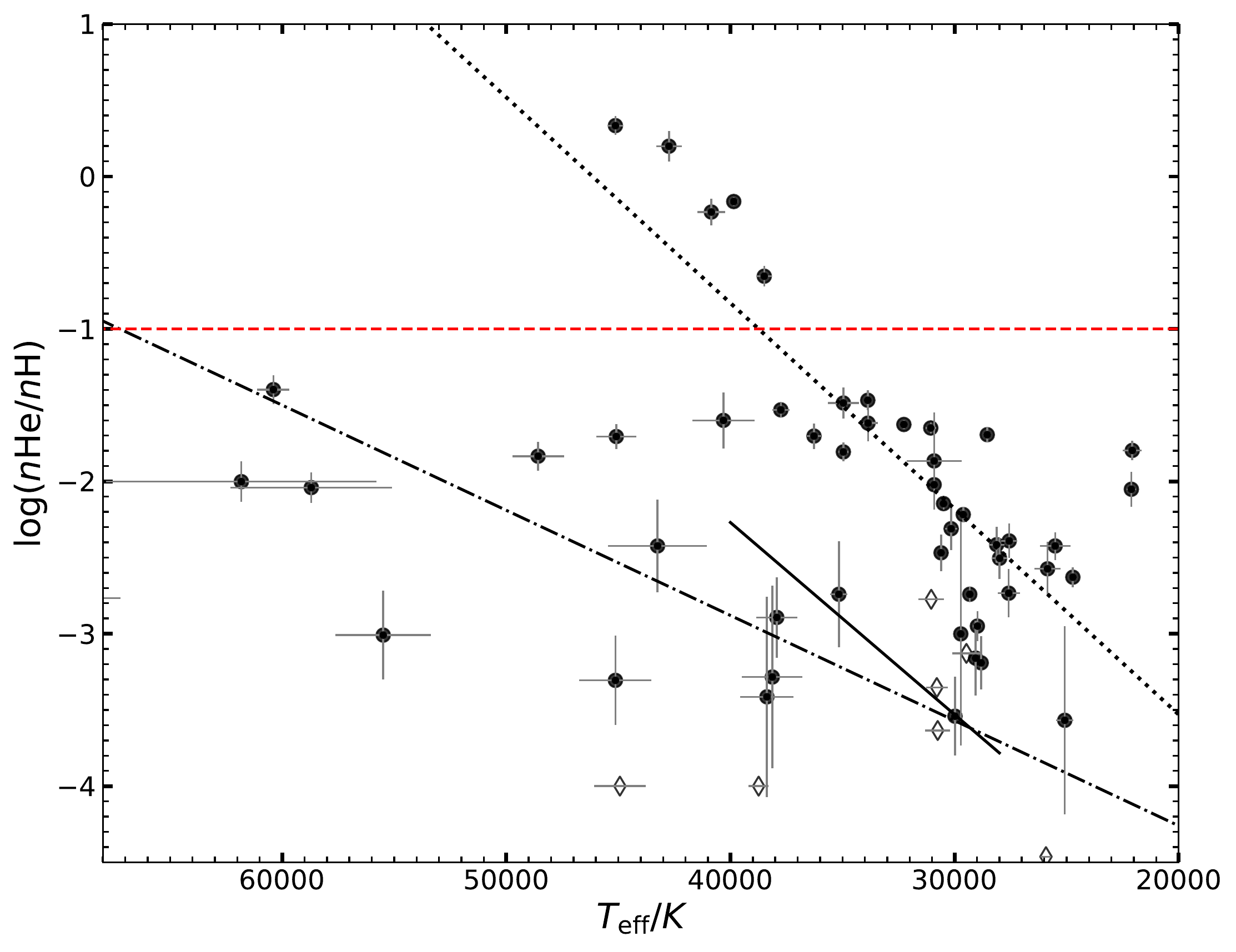}

\caption{$T_\mathrm{eff}$-$\mathrm{log}(n\mathrm{He}/n\mathrm{H})$ diagram  
for the 56 hot subdwarf stars identified in this study. The red dashed line denotes 
the solar He abundance. The dotted line and solid line are the linear regression line 
fitted by Edelmann et al. (2003), while the dot-dashed line is the best-fitting line 
for the He-poor sequence in N\'emeth et al. (2012). Diamonds denote the 
stars for which we just obtained  the upper limit 
of $\mathrm{log}(n\mathrm{He}/n\mathrm{H})$ (see Table 1). }
\end{figure}

Fig 7 shows the $T_\mathrm{eff}$-$\mathrm{log}(n\mathrm{He}/n\mathrm{H})$ 
diagram for our hot subdwarf stars.  The solar He abundance 
is marked by a horizontal red dashed line. The diamonds 
represent the stars for which only an upper limit of $\mathrm{log}(n\mathrm{He}/n\mathrm{H})$ 
could be obtained. Edelmann et al. (2003) found two He 
sequences, which are positive correlations between the effective temperature and He abundance 
(i.e., a He-rich sequence and a He-weak sequence) 
when the analyzed spectra of hot subdwarf stars were from the Hamburg Quasar Survey.  
The He-rich sequence of their sample follows the fitting formula:  
\begin{equation}
\mathrm{log}(n\mathrm{He}/n\mathrm{H})=-3.53+1.35\left(\frac{T_\mathrm{eff}}{10^{4}K}-2.00\right),  
\end{equation}
while the He-weak sequence in their study follows the formula:  
\begin{equation} 
\mathrm{log}(n\mathrm{He}/n\mathrm{H})=-4.79+1.26\left(\frac{T_\mathrm{eff}}{10^{4}K}-2.00\right).  
\end{equation}
These two lines  are shown by the dotted and the solid lines in Fig 7, respectively. 
We found results similar to those described by Edelmann et al. (2003), 
the two He sequences of hot subdwrf stars are also present in our sample. 
Moreover, the  He-rich sequence in Fig 7 
could be fitted well by the line described in equation (5), which is 
from Edelmann et al. (2003). However, a He-weak sequence in our sample  
follows a different trend than the He-weak sequence 
by Edelmann et al. (2003). On the other hand, the He-weak sequence in our sample   
is consistent with the one presented in N\'emeth et al. (2012). 
They used another line to fit the He-weak sequence in their study: 
\begin{equation}
\mathrm{log}(n\mathrm{He}/n\mathrm{H})=-4.26+0.69\left(\frac{T_\mathrm{eff}}{10^{4}K}-2.00\right). 
\end{equation}
We also plot the linear regression by equation (7), which 
is denoted by a dot-dashed line in Fig 7. 
The trend of this line is consistent with our He-weak sequence. Furthermore, 
Edelmann et al. (2003) also found two less clear sequences of hot subdwarf stars in the  
$\mathrm{log}$ \textit{g}-$\mathrm{log}(n\mathrm{He}/n\mathrm{H})$ plane. However, 
we did not find a similar result in our sample (see Fig 8). 

\begin{figure}
\centering
\includegraphics [width=100mm]{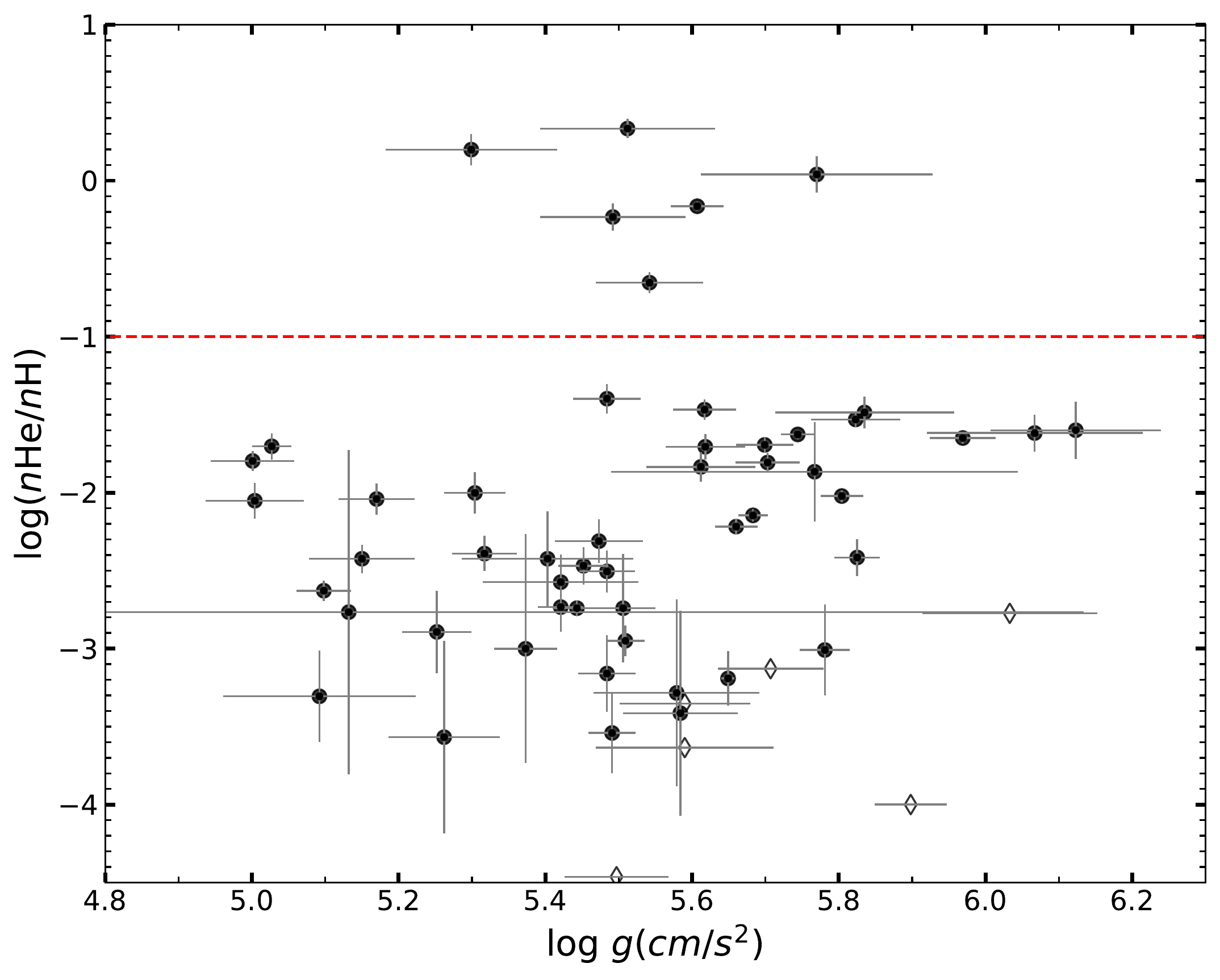}

\caption{The $\log{g}-\log{(n{\rm He}/n{\rm H})}$ plane 
for the 56 hot subdwarf stars identified in this study, 
the red dashed line marks the solar He abundance for reference. 
Diamonds denote the stars for which we just obtained  the upper limit 
of $\mathrm{log}(n\mathrm{He}/n\mathrm{H})$ (see Table 1).}
\end{figure}

\section{Discussion}
%\subsection{A new method to search for hot subdwarf stars in large spectroscopic surveys}
The traditional method to search for hot subdwarf stars in large 
spectroscopic surveys is to make color cuts followed by visual inspections.  This 
method requires homogeneous photometric information  to 
obtain the colors of the stars (e.g., \textit{u-g} and \textit{g-r}; Geier et al. 2011). 
Therefore, the traditional method is not  suitable for large spectral   
databases without supplementary photometric measurements, 
such as the spectral database of LAMOST.  
The HELM algorithm, as described in Paper I and in this study, 
does not need color information to filter out spectra with certain spectral properties. 
This makes HELM a suitable method to screen large spectroscopic surveys 
for hot subdwarf stars, or any other objects with distinct spectral features.

One may note that  He-rich hot subdwarf stars  
are under-represented  in our samples (e.g., only 5 stars with 
$\mathrm{log}(n\mathrm{He}/n\mathrm{H})>-1.0$, see Fig 7 in this paper), 
this could be due to the fact that the number of He-rich hot subdwarf stars 
in the training sample is small.  Our training spectra were the hot subdwarfs from Luo et al. (2016), which 
consists of 77 sdB stars, 12 sdO stars, 10 He-sdB stars and 15 He-sdO stars. 
According to the classification scheme of Luo et al. (2016), 
both sdB and sdO stars are He-poor hot subdwarf stars with dominant 
H Balmer lines, while both He-sdB and He-sdO stars are He-rich stars 
with dominant He I or He II lines. That is, there are many more 
hot subdwarf stars  with dominant H Balmer lines (He-poor stars) 
than the stars with dominant He lines (He-rich stars) in our training sample,   e.g., 77 versus 25. 
In addition to this, we did not separate these different type of subdwarf stars during the experiments.  
Instead, we trained HELM with all the sample spectra together,  
thus the larger the number of stars of a particular type in the training sample, 
the greater the precision with which this stellar type may be identified
in the science sample.  These factors could be accounted for the lack of 
He-rich hot subdwarf stars in our results. 
 
The quantity and quality of the training spectra are both very important factors in the 
HELM algorithm method, and have a direct influence on the results (Tang et al. 2015). 
Before we started this work, only 166 hot subdwarf stars (including 122 
single-lined stars and 44 double-lined stars) with LAMOST spectra  were 
published in Luo et al. (2016). Therefore, the number of 
hot subwarf stars is limited in our training spectra. Moreover, 
among 122 single-lined hot subdwarf stars, 8 stars are classified as BHB stars 
in Luo et al. (2016), and  only about 50 have a 
SNR larger than 10. As a result, although the initial 
candidates selected by HELM algorithm contains more than 7000 spectra, 
but nearly 6000 spectra have a $u$-band SNR below 10, which demonstrates  
a poor quality of the spectra for a follow-up  study.  These spectra have been 
discarded  from our analysis  as we mention in Section 3. 
With these considerations the total number 
of hot subdarfs in the LAMOST target list is likely much higher.  

Having used machine learning tools to search for  
hot subdwarf stars in LAMOST, we can outline some future improvements 
that will be required for a better efficiency and accuracy of the method.  For example, 
we plan to add the standard hot subdwarf stars listed in 
Drilling et al. (2013) into our training sample,  since it 
provides detailed classification for all kinds of hot 
subdwarf stars with different types, which will be 
quite useful to classify hot subdwarf stars by 
the HELM algorithm.  We also plan to 
cross match the LAMOST database with the newest 
hot subdwarf catalogue (e.g., Geier et al. 2017), then we 
will be able to add many high quality hot subdwarf spectra 
to our training sample. 
From these improvements we expect a large number of new subdwarfs 
to be uncoverd from the LAMOST survey in the near future.   These works are 
already on the way and will make important contributions 
on the study of the formation and evolution of hot subdwarf stars. 

\section{Summary}
We have applied the HELM algorithm in our study  to search for hot subdwarf stars in LAMOST DR1.  
56 hot subdwarf stars are identified among 465 candidates with single-lined  
spectra, and their atmospheric parameters have been obtained by fitting the profiles of 
H Balmer and He lines with the synthetic spectra calculated from NLTE model 
atmospheres.  31 sdB stars, 11 sdO stars, 9 sdOB stars,   
4 He-sdOB and 1 He-sdO stars were found in our study. These stars confirm the two 
He sequences of hot subdwarf stars in $T_\mathrm{eff}$-$\log{(n{\rm He}/n{\rm H})}$ diagram,  
which were first found by Edelmann et al. (2003). 

Our study has shown the strength of the HELM algorithm to filter out targets 
with specific spectral properties from large sets of spectroscopic data directly, 
without the need of any photometric observations or pre-selection.  
Though the total number of hot subdwarf stars identified   
may seem low compared to the sample size,  
it is mainly due to the limited quantity and quality of the training spectra.  
We expect that many more hot subdwarf stars will be  found in the 
LAMOST database using machine learning method in the future after 
our experiences are implemented in the algorithm. 
We used the HELM algorithm for the first time to search for 
hot subdwarf stars in a large spectroscopic survey, and the results presented in our study demonstrate that  this method could be applied to search for other types of object with obvious features in their spectra or images.  

%%%%%%%%%%%%%%%%%%%%%%%%%%%%%%%%%%%%%%%

\begin{ack}
We thank the referee, A. E. Lynas-Gray, for his valuable suggestions and comments, which improved the manuscript much. 
This work was supported by the National Natural Science Foundation
of China Grant Nos, 11390371, 11503016, 11873037，11603012 and U1731111,    
Natural Science Foundation of Hunan province Grant No. 2017JJ3283,  
the Youth Fund project of Hunan Provincial Education Department
Grant No. 15B214, the Astronomical Big Data Joint Research
Center, co-founded by the National Astronomical Observatories, Chinese
Academy of Sciences and the Alibaba Cloud, Young Scholars Program of Shandong University, 
Weihai 2016WHWLJH09, Natural Science Foundation of Shandong Province ZR2015AQ011, China post-doctoral 
Science Foundation 2015M571124. This research has used the services of \mbox{www.Astroserver.org} under reference W00QEL. P.N. acknowledges support from the Grant Agency of the Czech Republic (GA\v{C}R 18-20083S). 
The LAMOST Fellowship is supported by Special Funding for Advanced Users, 
budgeted and administered by the Center for Astronomical 
Mega-Science, Chinese Academy of Sciences (CAMS). 
Guoshoujing Telescope (the Large Sky Area Multi-Object Fiber 
Spectroscopic Telescope LAMOST) is a National Major Scientific 
Project built by the Chinese Academy of Sciences. 
Funding for the project has been provided by the 
National Development and Reform Commission. 
LAMOST is operated and managed by the National Astronomical Observatories, 
Chinese Academy of Sciences.
\end{ack}

%%%
% See the manual for the detail.
%%%


\begin{thebibliography}{}
% Journals(e.g. A\&A,ApJ,AJ,NMRAS,PASP ...)
% Authors, Year, Journal, Vol#, Page#
% Journal Title Abbreviation >> http://www.asj.or.jp/pasj/Jabb.html
\bibitem[Beers et al.(1992)]{1992AJ....103..267B} Beers, Timothy C., Preston, George W., Shectman, Stephen A.,  et al.\ 1992, \aj, 103, 267
\bibitem[Brown et al.(2016)]{2016ApJ...822...44B} Brown, T. M., Cassisi, S., D'Antona, F.,  et al.\ 2016, \apj, 822, 44
\bibitem[Bu et al.(2017)]{2017ApJS..233....2B} Bu, Yude., Lei, Zhenxin., Zhao, Gang.,  et al.\ 2017, \apjs, 233, 2 (Paper I)
\bibitem[Catelan(2009)]{2009Ap&SS.320..261C} Catelan, M.\ 2009, \apss, 320, 261
\bibitem[Chen et al.(2013)]{2013MNRAS.434..186C} Chen, Xuefei., Han, Zhanwen., Deca, Jan.,  et al.\ 2013, \mnras, 434, 186
\bibitem[Clewley et al.(2002)]{2002MNRAS.337...87C} Clewley, L., Warren, S. J., Hewett, P. C., et al. \ 2002, \mnras, 337, 87
\bibitem[Copperwheat et al.(2011)]{2011MNRAS.415.1381C} Copperwheat, C. M., Morales-Rueda, L., Marsh, T. R.,  et al.\ 2011, \mnras, 415, 1381
\bibitem[Cui et al.(2012)]{2012RAA....12.1197C} Cui, Xiang-Qun., Zhao, Yong-Heng., Chu, Yao-Quan., et al.\ 2012, RAA, 12, 1197
\bibitem[Dorman et al.(1993)]{1993ApJ...419..596D} Dorman, Ben., Rood, Robert T., \& O'Connell, Robert W.\ 1993, \apj, 419, 596
\bibitem[Drilling et al.(2013)]{2013A&A...551A..31D} Drilling, J. S., Jeffery, C. S., Heber, U.,   et al.\ 2013, \aap, 551, 31
\bibitem[Edelmann et al.(2003)]{2003A&A...400..939E} Edelmann, H., Heber, U., Hagen, H.-J.,  et al.\ 2003, \aap, 400, 939
\bibitem[Eisenstein et al.(2006)]{2006ApJS..167...40E} Eisenstein, Daniel J., Liebert, James., Harris, Hugh C., et al.\ 2006, \apjs, 167, 40
\bibitem[Gaia Collaboration et al.(2018)]{} Gaia Collaboration, Brown, A., Vallenari, A., et al. 2018a,  arXiv:1804.09365
\bibitem[Gaia Collaboration et al.(2018)]{} Gaia Collaboration, Babusiaux, C., van Leeuwen, F., et al. et al. 2018b,  arXiv:1804.09378
\bibitem[Geier et al.(2011)]{2011A&A...530A..28G} Geier, S., Hirsch, H., Tillich, A., et al.\ 2011, \aap, 530, 28
%\bibitem[Geier et al.(2013)]{2013A&A...557A.122G} Geier, S., Heber, U., Edelmann, H., et al.\ 2013, \aap, 557, 122
\bibitem[Geier et al.(2015)]{2015Sci...347.1126G} Geier, S., Fürst, F., Ziegerer, E., et al.\ 2015, Science, 347, 1126
\bibitem[Geier et al.(2017)]{2017A&A...600A..50G} Geier, S., $\O$stensen, R. H., N\'emeth, P., et al.\ 2017, \aap, 600, 50 
\bibitem[Greenstein \& Sargent (1974)]{1974ApJS...28..157G} Greenstein, Jesse L., \& Sargent, Anneila I. \ 1974, \apjs, 28, 157 
\bibitem[Han et al.(2002)]{2002MNRAS.336..449H} Han, Z., Podsiadlowski, Ph., Maxted, P. F. L., et al.\ 2002, \mnras, 336, 449
\bibitem[Han et al.(2003)]{2003MNRAS.341..669H} Han, Z., Podsiadlowski, Ph., Maxted, P. F. L., et al. \ 2003, \mnras, 341, 669
\bibitem[Han et al.(2007)]{2007MNRAS.380.1098H} Han, Z., Podsiadlowski, Ph., \& Lynas-Gray, A. E. \ 2007, \mnras, 380, 1098
\bibitem[Heber(1987)]{1987MitAG..70...79H} Heber, U.\ 1987, MitAG, 70, 79
\bibitem[Heber(2009)]{2009ARA&A..47..211H} Heber, U.\ 2009, \araa, 47, 211
\bibitem[Heber(2016)]{2016PASP..128h2001H} Heber, U.\ 2016, \pasp, 128, 2001
\bibitem[Huang et al.(2006)] {} Huang, G., Zhu, Q., \& Siew, C.\ 2006,  Neurocomputing, 70, 489
\bibitem[Hubeny \& Lanz(1995)]{1995ApJ...439..875H} Hubeny, I., Lanz, T. \ 1995, \apj, 439, 875 
\bibitem[Hubeny \& Lanz(2017)]{2017arXiv170601859H} Hubeny, I., \& Lanz, T.\ 2017, arXiv:1706.01859 
\bibitem[Kawka et al.(2015)]{2015MNRAS.450.3514K} Kawka, A., Vennes, S., O'Toole, S., et al.\ 2015, \mnras, 450, 3514 
\bibitem[Kepler et al.(2015)]{2015MNRAS.446.4078K} Kepler, S. O., Pelisoli, I., Koester, D., et al.\ 2015, \mnras, 446, 4078
\bibitem[Kepler et al.(2016)]{2016MNRAS.455.3413K} Kepler, S. O., Pelisoli, I., Koester, D.,  et al.\ 2016, \mnras, 455, 3413
\bibitem[Lanz \& Hubeny(1995)]{1995ApJ...439..905L} 	Lanz, T., Hubeny, I. \ 1995, \apj, 439, 905 
\bibitem[Lanz \& Hubeny(2007)]{2007ApJS..169...83L} 	Lanz, T., Hubeny, I. \ 2007, \apjs, 169, 83 
%\bibitem[Lanz et al.(2004)]{2004ApJ...602..342L} Lanz, Thierry., Brown, Thomas M., Sweigart, Allen V., et al.\ 2004, \apj, 602, 342
\bibitem[Lei et al.(2015)]{2015MNRAS.449.2741L} Lei, Zhenxin., Chen, Xuemei., Zhang, Fenghui., et al.\ 2015, \mnras, 449, 2741
\bibitem[Lei et al.(2016)]{2016MNRAS.463.3449L} Lei, Zhenxin., Zhao, Gang., Zeng, Aihua.,  et al.\ 2016, \mnras, 463, 3449
\bibitem[Li et al.(2015)] {} Li, J., Du, Q., Li, W., et al.\ 2015,  JARS, 9, 097296
\bibitem[Luo et al.(2015)]{ 2015RAA....15.1095L} Luo, A.-Li., Zhao, Yong-Heng., Zhao, Gang.,  et al.\ 2015, RAA, 15, 1095 
\bibitem[Luo et al.(2016)]{2016ApJ...818..202L} Luo, Yang-Ping., N\'emeth, P., Liu, Chao.,  et al.\ 2016, \apj, 818, 202
\bibitem[Mao et al.(2014)] {} Mao, L., Zhang, L., Liu, X., et al.\ 2014,  Mathematical Problems in Engineering, 2014, 1
\bibitem[Maxted et al.(2001)]{2001MNRAS.326.1391M} Maxted, P. F. L., Heber, U., Marsh, T. R., et al.\ 2001, \mnras, 326, 139 
%\bibitem[Michaud et al.(2011)]{2011A&A...529A..60M} Michaud, G., Richer, J., Richard, O.\ 2011, \aap, 529, 60
%\bibitem[Miller Bertolami et al.(2008)]{2008A&A...491..253M} Miller Bertolami, M. M., Althaus, L. G., Unglaub, K., et al.\ 2008, \aap, 491, 253 
\bibitem[Minhas et al.(2010)] {} Minhas, R., Baradarani, A., Seifzadeh,S., et al.\ 2010, Neurocomputing, 73, 1906
\bibitem[Moehler et al.(1990)]{1990A&AS...86...53M} Moehler, S., Richtler, T., de Boer, K. S.,  et al.\ 1990, \aaps, 86, 53
\bibitem[Napiwotzki et al.(2004)]{2004Ap&SS.291..321N} Napiwotzki, R., Karl, C. A., Lisker, T.,  et al.\ 2001, \apss, 291, 321
\bibitem[N\'emeth et al.(2014)]{ 2014ASPC..481...95N} N\'emeth, P., $\O$stensen, R., Tremblay, P., et al.\ 2014, ASPC, 481, 95
\bibitem[N\'emeth(2012)]{2012MNRAS.427.2180N} N\'emeth, P., Kawka, A., \&  Vennes, S.\ 2012, \mnras, 427, 2180
\bibitem[O'Connell(1999)]{1999ARA&A..37..603O} O'Connell, Robert W. \ 1999, \araa, 37, 603 
\bibitem[$\O$stensen et al.(2006)]{2006BaltA..15...85O} $\O$stensen, R. H.\ 2006, Baltic, 15, 85 
\bibitem[$\O$stensen et al.(2010)]{2010MNRAS.409.1470O} $\O$stensen, R. H., Silvotti, R., Charpinet, S.,  et al.\ 2010, \mnras, 409, 1740 
\bibitem[Paczy\'nski(1971)]{1971AcA....21....1P} Paczy\'nski, B.\ 1971, Acta Astron, 21, 1
\bibitem[S\'ersic(1968)]{1968adga.book.....S} S\'ersic, J.~L.\ 1968, Cordoba, Argentina: Observatorio Astronomico, 1968
\bibitem[Sirko et al.(2004)]{2004AJ....127..899S} Sirko, Edwin., Goodman, Jeremy., Knapp, Gillian R., et al.\ 2004, \aj, 127, 899
\bibitem[Tang et al.(2015)] {} Tang, J., Deng, C., \& Huang, G. \ 2015,  ITNN, 27, 809
\bibitem[Tremblay \& Bergeron (2009)] {2009ApJ...696.1755T} Tremblay, P.-E., Bergeron, P. \ 2015,  \apj, 696, 1755 
\bibitem[Vennes, Kawka \& Németh(2011)]{2011MNRAS.410.2095V} Vennes, S., Kawka, A \& Németh, P. \ 2011, \mnras, 410, 2095
\bibitem[Wang et al.(2009)]{2009MNRAS.395..847W} Wang, B., Meng, X., Chen, X., et al.\ 2009, \mnras, 395, 847
\bibitem[Xiong et al.(2017)]{2017A&A...599A..54X} Xiong, H., Chen, X., Podsiadlowski, Ph.,  et al.\ 2017, \aap, 599, 54
\bibitem[Xue et al.(2008)]{2008ApJ...684.1143X} Xue, X. X., Rix, H. W., Zhao, G., et al.\ 2008, \apj, 684, 1143
\bibitem[York et al.(2000)]{2000AJ....120.1579Y} York, Donald G., Adelman, J., Anderson, John E., et al.\ 2000, \aj, 120, 1579
\bibitem[Zhang \& Jeffery(2012)]{2012MNRAS.419..452Z} Zhang, Xianfei., \&  Jeffery, C. S.\ 2012, \mnras, 419, 452
\bibitem[Zhang et al.(2017)]{2017ApJ...835..242Z} Zhang, Xianfei., Hall, Philip D.., Jeffery, C. Simon., et al.\ 2017, \apj, 835, 242
\bibitem[Zhao et al.(2006)]{2006ChJAA...6..265Z} 	Zhao, Gang., Chen, Yu-Qin., Shi, Jian-Rong., et al.\ 2006, ChJAA, 6, 265  
\bibitem[Zhao et al.(2012)]{2012RAA....12..723Z} Zhao, Gang., Zhao, Yong-Heng., Chu, Yao-Quan., et al.\ 2012, RAA, 12, 723 
\end{thebibliography}
\end{document}